\documentclass{aa}     
\usepackage{epsf}
\usepackage{amssymb}
\usepackage{amsmath}
\usepackage{amsxtra}
\usepackage{times}
\usepackage{graphicx}
 
\def\aa{A\&A}

\begin{document}

   \title{The Double Quasar HE~1104$-$1805: a case study for time delay 
   determination with poorly sampled lightcurves}

   \author{Rodrigo Gil-Merino, Lutz Wisotzki, Joachim Wambsganss}
 
   \institute{Universit\"at Potsdam, 
              Am Neuen Palais 10, D-14469 Potsdam, Germany\\
		  \email{rmerino, lutz, jkw@astro.physik.uni-potsdam.de}}
		  
   \offprints{R. Gil-Merino}
 
\date{Received date; accepted date}
\date{Submitted: July, 2001}
 
\titlerunning{HE~1104$-$1805: time delay with poorly sampled data}
\authorrunning{R. Gil-Merino, L. Wisotzki \& J. Wambsganss}
 
\abstract{We present a new determination of the time 
delay of the gravitational lens system HE~1104$-$1805 ('Double Hamburger') 
based on a previously unpublished dataset. We argue that 
the previously published value of $\Delta t_{A-B}=0.73$ years was affected by a
bias of the employed method. We determine a new value of 
$\Delta t_{A-B}=0.85\pm0.05$ years 
(2$\sigma$ confidence level), using six different techniques 
based on non interpolation methods in the time domain. The result demonstrates 
that even in the case of poorly sampled lightcurves, useful information can be 
obtained with regard to the time delay. The error estimates were calculated 
through Monte Carlo simulations. 
With two already existing models for the lens and using its recently determined 
redshift, we infer a range of values of the Hubble parameter: 
$H_0=48\pm4~\mathrm{km}~\mathrm{s}^{-1}~\mathrm{Mpc}^{-1}$ (2$\sigma$)
for a singular isothermal ellipsoid (SIE) and 
$H_0=62\pm4~\mathrm{km}~\mathrm{s}^{-1}~\mathrm{Mpc}^{-1}$ (2$\sigma$)
for a constant mass-to-light ratio plus shear model ($M/L+\gamma$). 
The possibly much larger errors due to systematic uncertainties in 
modeling the lens potential are not included in this error estimate.
}

\maketitle

\section{Introduction}
The double quasar HE~1104$-$1805 at a redshift of $z_Q  = 2.319$ was 
originally discovered  by Wisotzki  et al. (1993). The two images
with (original) B magnitudes of 16.70 and 18.64 are separated by
$3\farcs 195$ (Kochanek et al. 1998). 
The spectral line ratios and profiles turned out to be almost identical 
between the two images, but image A has a distinctly harder continuum.
Wisotzki et al. (1995) report about a dimming of both components over 
about 20 months, accompanied by a softening of the
continuum slope of both images.
The lensing galaxy was discovered by Courbin et al. (1998) in the NIR 
and by Remy et al. (1998) with HST. The authors tentatively identified 
the lensing galaxy with a previously detected damped Lyman alpha system 
at $z=1.66$ (Wisotzki et al. 1993; 
Smette et al. 1995; Lopez et al. 1999). This identification, however, 
was disputed by Wisotzki et al. (1998).
Using FORS2 at the VLT, Lidman et al. (2000) finally 
determined the redshift of the lensing galaxy  to $z_G = 0.729\pm0.001$.

A first determination of the time delay in this system
was published by Wisotzki et al. (1998), based on five years of
spectrophotometric monitoring of HE1104~$-$1805, in which the
quasar images varied significantly, while the 
emission line fluxes appear to have remained constant. 
The Wisotzki et al. (1998) value for the time delay was 
$\Delta t_{A-B}=0.73$ years (no formal error
bars were reported), but they cautioned that a value as small as 0.3 years
could not be excluded.

HE~1104$-$1805 shows strong and clear
indications of gravitational microlensing, in particular based
on the continuum variability  with the line fluxes
almost unaffected (Wisotzki et al. 1993, Courbin et al. 2000).

Here we present an analysis of previously unpublished photometric
monitoring data of HE~1104$-$1805. First the data and light curves are 
presented (Sect. 2), 
then a number of numerically techniques are 
described and discussed and, as the scope of this paper is a comparison of different
techniques in the case of poorly sampled data, we finally
applied to this data set, in order to determine
the time delay (Sect. 3).  
A discussion of the results and the implications for the value of the 
Hubble constant based on this new
value of the time delay and on previously avalaible lens models
are given in Sect. 4. 
In Sect. 5 we present our conclusions.

\section{Data acquisition and reduction}
\begin{figure}[hbtp]
 \centering
 \includegraphics[bbllx=62,bblly=78,bburx=353,bbury=439,width=8.5cm,
                  clip=true]{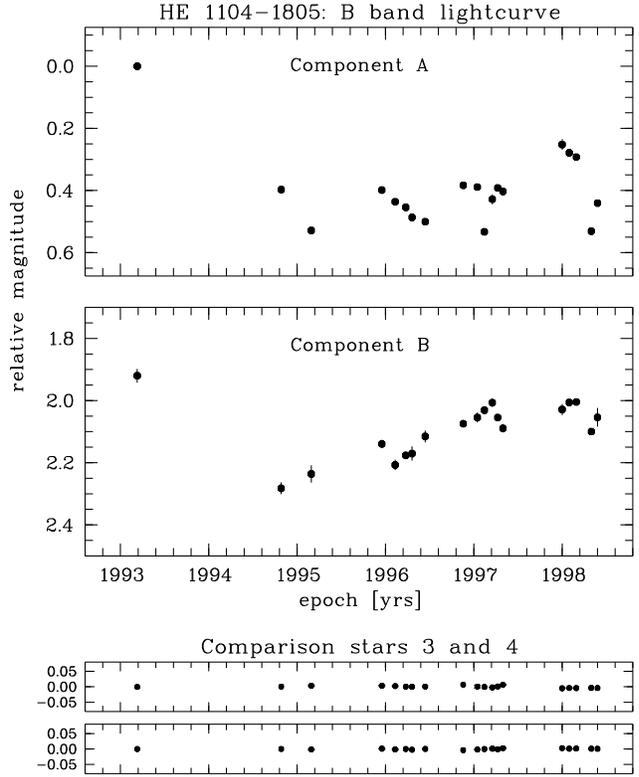}
 \caption[]{The new photometric dataset running from 1993 to 1998. The 
  zero point for the relative photometry of HE~1104$-$1805 is the first data
  point of component A (see Table \ref{Table1} for error estimates).}
 \label{Fig1} 
\end{figure}

Between 1993 and 1998, we obtained a $B$ band lightcurve of HE~1104$-$1805 
at 19 independent epochs, mostly in the course of a monitoring 
campaign conducted at the ESO 3.6\,m telescope in service
mode. The main intention of the programme was to follow the spectral
variations by means of relative spectrophotometry, but at each
occasion also at least one frame in the $B$ band was taken. 
A \emph{continuum} lightcurve, derived from the spectrophotometry, 
and a first estimate of the time delay were presented by \cite{W98} (1998; 
hereafter W98); 
details of the monitoring will be given in a forthcoming paper 
(Wisotzki \& L\'opez, in preparation). Here we concentrate on the broad band 
photometric data which have not been published to date.
Images were taken typically once a month during the visibility period.
The instrument used was EFOSC1 with 512$\times$512 pixels Tektronix 
CCD until June 1997, and EFOSC2 with a 2K$\times$2K Loral/Lesser
chip afterwards. The $B$ band frames (which were also used as
acquisition images for the spectroscopy) were always exposed for 
30 seconds, which ensured that also the main comparison stars were 
unsaturated at the best recorded seeing of $1\arcsec$. Sometimes
more than one exposure was made, enabling us to make independent
estimates of the photometric uncertainties. A journal of the
observations is presented together with the measured lightcurve
in Table \ref{Table1}.
The CCD frames were reduced in a homogeneous way following
standard procedures. After debiasing and flatfielding, photometry
of all sources in the field was conducted using the DAOPHOT~II
package (\cite{Stetson87}) as implemented into ESO-MIDAS.
The instrumental magnitudes of the QSO components and 
reference stars 1--5 (following the nomenclature of \cite{W95}) 
were recorded and placed on 
a homogeneous relative magnitude scale defined by the
variance-weighted averages over all comparison stars.
In Fig. \ref{Fig1} we show the resulting QSO lightcurves,
together with the two brightest comparison stars. The variability
of both QSO components is highly significant, including strong
fluctuations on the barely sampled timescales of months. This behaviour 
is stronger in component A, while component B leads the variability. The error 
estimates include shot noise, PSF fitting uncertainties and standard 
deviations in case of multiple images at a given epoch. Note the
similarity of these $B$ band data with the completely
independently calibrated continuum lightcurves of W98.
\begin{table}[tb]
\centering
\begin{tabular}{lllll}
\hline\noalign{\smallskip}
Epoch [yrs] & $\Delta B_A$ & $\sigma_{B_A}$ & $\Delta B_B$ & $\sigma_{B_B}$\\
\noalign{\smallskip}\hline\noalign{\smallskip}
 1993.19 & 0.000 & 0.009 & 1.920 & 0.022 \\
 1994.82 & 0.397 & 0.009 & 2.282 & 0.019 \\
 1995.16 & 0.529 & 0.008 & 2.236 & 0.028 \\
 1995.96 & 0.399 & 0.012 & 2.140 & 0.014 \\
 1996.11 & 0.436 & 0.008 & 2.207 & 0.017 \\
 1996.23 & 0.454 & 0.005 & 2.176 & 0.013 \\
 1996.30 & 0.486 & 0.009 & 2.171 & 0.023 \\
 1996.45 & 0.500 & 0.008 & 2.115 & 0.019 \\
 1996.88 & 0.383 & 0.007 & 2.074 & 0.012 \\
 1997.04 & 0.389 & 0.007 & 2.054 & 0.016 \\
 1997.12 & 0.533 & 0.009 & 2.031 & 0.013 \\
 1997.21 & 0.428 & 0.016 & 2.007 & 0.015 \\
 1997.27 & 0.392 & 0.007 & 2.055 & 0.012 \\
 1997.33 & 0.403 & 0.008 & 2.089 & 0.014 \\
 1998.00 & 0.252 & 0.017 & 2.029 & 0.018 \\
 1998.08 & 0.279 & 0.004 & 2.006 & 0.011 \\
 1998.16 & 0.292 & 0.004 & 2.004 & 0.011 \\
 1998.33 & 0.531 & 0.006 & 2.100 & 0.011 \\
 1998.40 & 0.441 & 0.007 & 2.054 & 0.030 \\
\noalign{\smallskip}\hline
\end{tabular}
\caption{$B$ band lightcurve data for HE~1104$-$1805. The first 
measurement of component A has arbitrarily been set to zero. The
error estimates include shot noise, PSF fitting uncertainties,
and also standard deviations in case of multiple images at a 
given epoch.}
\label{Table1}
\end{table}
\section{Time Delay Determination}

\subsection{Dispersion spectra method}
A first estimation for the time delay in this system resulted in a value of 
$\Delta t_{B-A}= -0.73$ years (W98), using the dispersion spectra method 
developed by Pelt et al. (1994, 1996; hereafter P94 and P96, respectly). Note
that we will express the time delay as $\Delta t_{B-A}$ (instead of 
$\Delta t_{A-B}$), since B leads the variability 
(see Fig. \ref{Fig1}), and thus there appears a minus sign in the result. 
We shall demonstrate below that the dispersion spectra method is not bias-free.
To facility a better understanding of this claim, we first briefly describe the
method in the following: The two time series $A_i$ and $B_j$ can
be expressed, using the P96 notation, as
\begin{eqnarray}
A_i=q(t_i)+\epsilon_A(t_i), i=1,...,N_A \\
B_j=q(t_j-\tau)+l(t_j)+\epsilon_B(t_j), j=1,...,N_B
\end{eqnarray}
$q(t)$ being the intrinsic variability of the quasar, $\tau$ the time delay and
$l(t)$ the magnification ratio plus another possible noise component (this 
could be pure noise or microlensing). Both lightcurves $A_i$ and $B_j$ are 
combined into a new one, $C_k$, for each value of the pair $(\tau, l(t))$, 
`correcting' $B_j$ serie by $l(t)$ in magnitudes and by $\tau$ in time
\begin{equation}
C_k(t_k)=\left \{\begin{array}{ll}
           A_i         & if\ t_k=t_i     \\
	   B_j-l(t_j)  & if\ t_k=t_j-\tau 
          \end{array}
	 \right.,
\end{equation} 
with $k=1,...,N$ and $N=N_A+N_B$. Then the dispersion spectrum is calculated 
analytically by the expression 
\begin{equation}
D^2_{4,k}=\min_{l(t)}\frac{\sum\limits^{N-1}_{n=1}\sum\limits^N_{m=n+1}
S^{(k)}_{n,m}W_{n,m}G_{n,m}(C_n-C_m)^2}{\sum\limits^{N-1}_{n=1}
\sum\limits^N_{m=n+1}S^{(k)}_{n,m}W_{n,m}G_{n,m}},
\end{equation}
where $W_{n,m}$ are the statistical weights; $G_{n,m}=1$ if the points $C_n$ 
and $C_m$ come from different time series, $A_i$ or $B_j$, and is $0$ otherwise;
and $S^{(k)}_{n,m}$ is a function that weights each difference $(C_n-C_m)$ 
depending on the distance between the points. In P96 they show three possible
definitions for this function, here we have selected
\begin{equation}
S^{(2)}_{n,m}=\left \{\begin{array}{ll}
  1-\frac{\left| t_n-t_m \right|}{\delta}  
                     & if\ \left|t_n-t_m\right| \leq \delta\\
  0  
                     & if\ \left|t_n-t_m\right|  > \delta 
  \end{array}
  \right.,
\end{equation}
which includes those pairs for which the distance between two observations is
less than a certain \emph{decorrelation length} $\delta$. 
More details can be found in P94 and P96.
The definition of this function here is slightly different from the one 
used in W98. We have two reasons to do so: first, we will demonstrate that 
the selection of one or another definition does not play a crucial 
role in this case; second, the function $S^{(3)}_{n,m}$ used in W98 is
supposed to avoid the problem of having big gaps between the observational 
points in the lightcurves, but we will try to solve this problem in a different
way.

The new dataset used here has the same sampling as the one used for the
first estimation of the time delay in W98. As the errorbars for individual points 
are also very similar, one should expect to obtain a similar time delay. And in 
fact this is exactly what happens when applying the dispersion spectra method as 
described above.
The original dataset is plotted in Fig. \ref{Fig1}. There are 19 observational
points for each component. We apply the dispersion spectra method (P94, P96): the 
result is $\Delta t_{B-A}=-0.73$ years, i.e., the same value as the first published 
estimation.
 
Since W98 did not provide a formal error estimate, we now investigate the goodness of this value and try to
estimate the uncertainty, and we also want to check 
the self-consistency of the method in this case. For this purpose we do a test based on 
an iterative procedure: after having applied the dispersion spectra method 
to the whole data set, 
we make a selection of the data trying to avoid big gaps between the 
epochs and considering points in both lightcurves that fall in the same time interval 
once one has corrected the time shift with the derived time delay. This will 
avoid the so-called border effects, and a time delay close 
to the initial one should result when the dispersion spectra are 
recalculated for the selected data. We do this in the next subsection.

\subsection{Borders and gaps}			
\label{borders}
\begin{figure}[hbtp]
 \centering
 \includegraphics[bbllx=60,bblly=102,bburx=360,bbury=369,width=8.5cm,
                  clip=true] {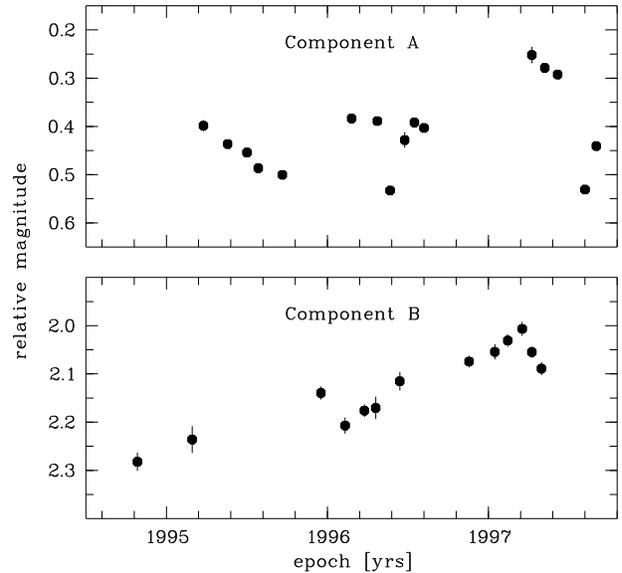}
 \caption[]{The first point of the whole dataset has been removed and then 
  the points that do not fall in the same time interval once we have shifted
  the A lightcurve with the value of the first time delay estimation, 
  $\Delta t_{B-A}=-0.73$ years. Thus component A has now 16 points and
  the component B 13 points. If the procedure were self-consistent and the first
  time delay estimation right, we would
  naturaly expect a confirmation of this value in a second measurement of the delay by
  using the new dataset.}
 \label{Fig2}
\end{figure}

We first consider $\Delta t_{B-A}=-0.73$ years as a 
first rough estimate of the time delay, in agreement with W98. 
It is obvious that using this time delay, the first 
point of the whole dataset (epoch 1993.19) of component B has no 
close partner in component A. Eliminating this point means avoiding 
the big gap of almost two years at the beginning of the lightcurves. Once 
this is done, the last five points of the lightcurve B and the first two ones of A
(after eliminating the epoch 1993.19) are not useful anymore for a time delay 
determination since they do not cover the same intrinsic time interval. 
We also eliminate these points. Now we 
have a `clean' dataset with 16 points from component A and 13 points from component 
B. The situation is illustrated in Fig. \ref{Fig2}, where only the epochs inside the 
time interval [1994.5, 1998.0] are plotted. This is the time interval for which the two 
lightcurves overlap after the $-0.73$ years correction for component A.

\begin{figure}[hbtp]
 \centering
 \epsfxsize=6 cm
 \rotatebox{-90}{\epsffile{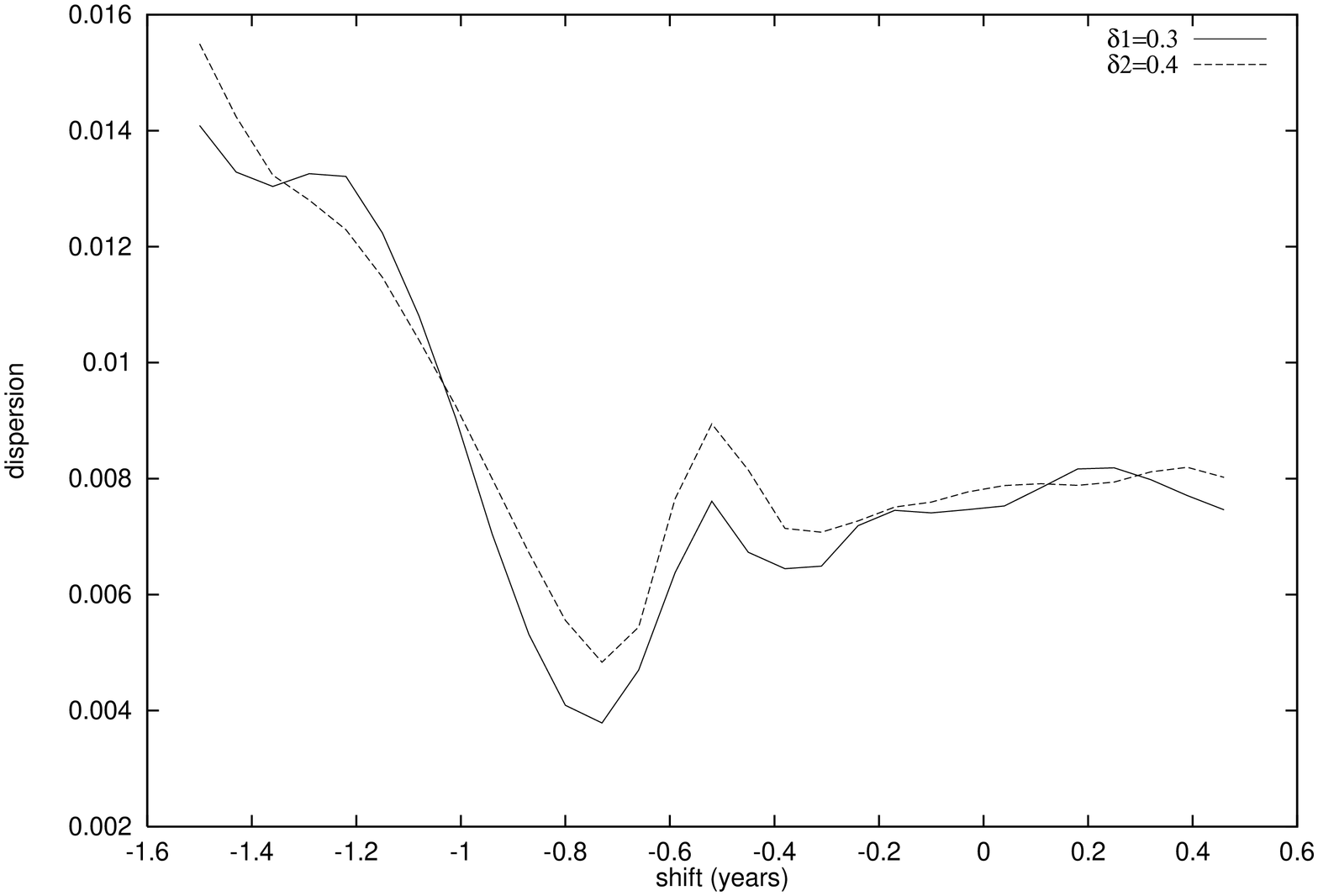}}
 \epsfxsize=6 cm
 \rotatebox{-90}{\epsffile{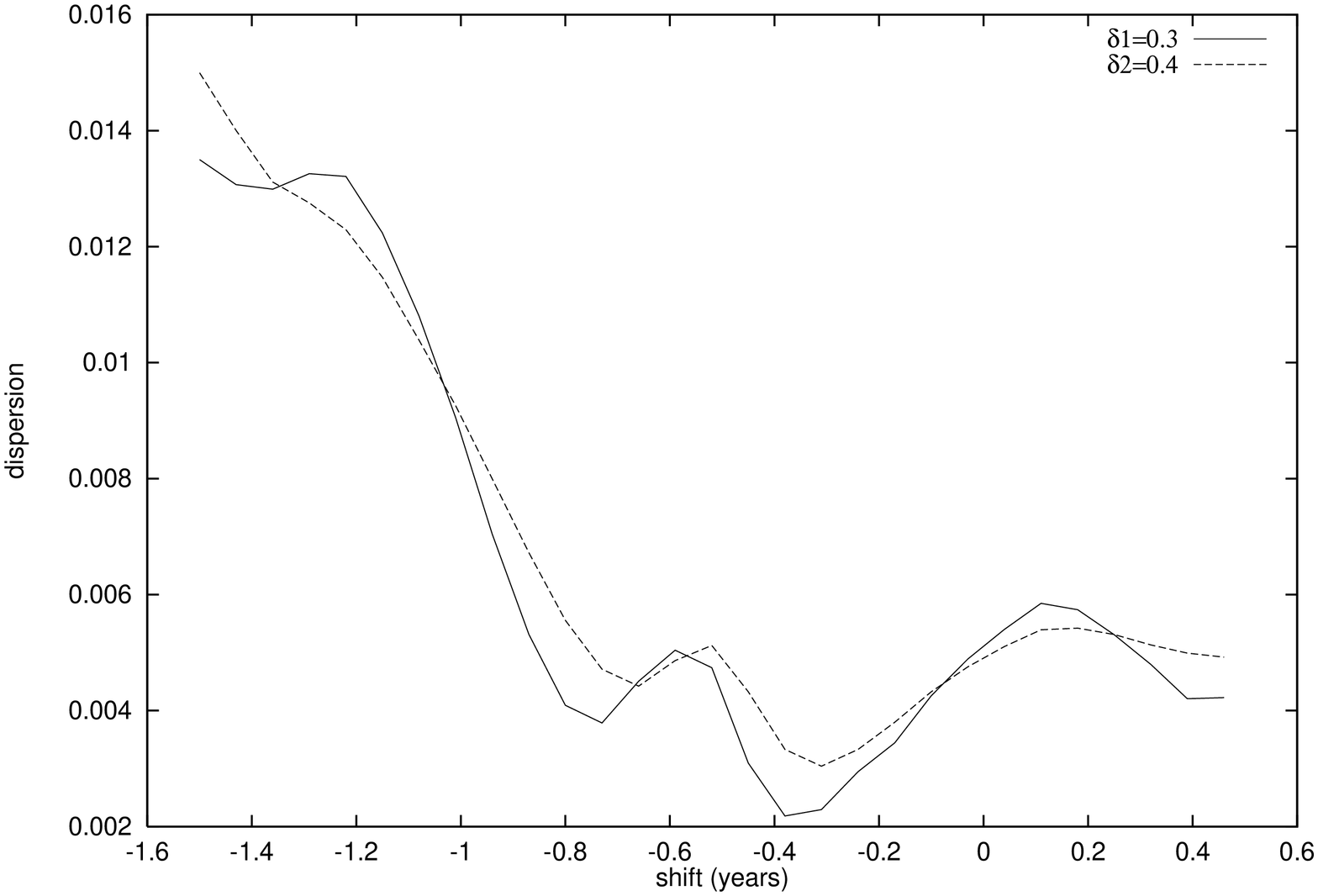}}
 \epsfxsize=6 cm
 \rotatebox{-90}{\epsffile{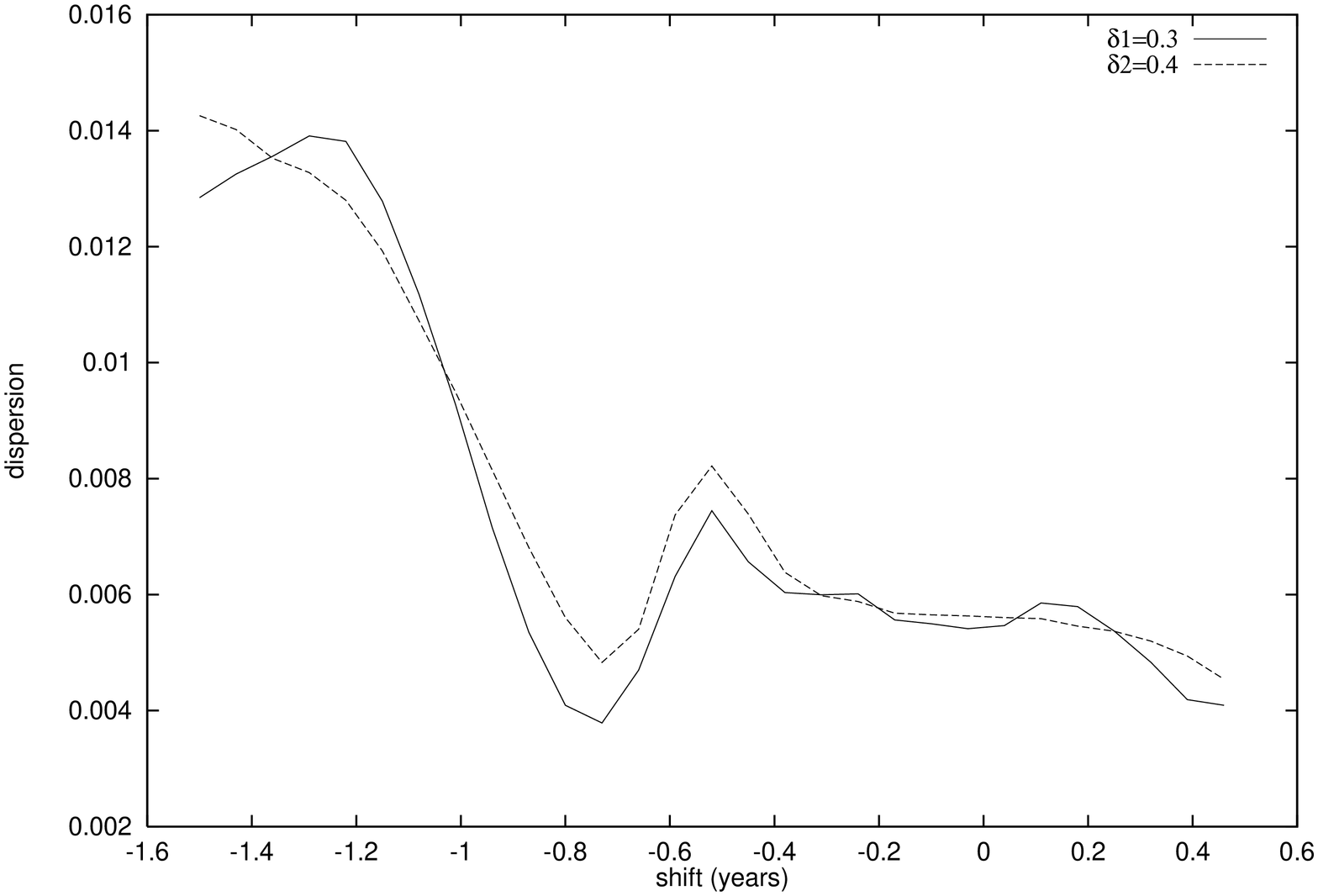}}
 \caption[]{Dispersion spectra: The \emph{upper panel} shows the result when 
 all the points are taken into account. In the \emph{middle panel}, the 
 result after correcting borders with the first estimation of the time delay, 
 i.e. $\Delta t_{B-A}=-0.73$ years. In the \emph{bottom panel} we use a correction of 
 $-0.38$ years obtained in the middle panel. We recover the previous
 value for the time delay of $\Delta t_{B-A}=-0.73$ years, showing the 
 inconsistency of the method.
 In each subfigure, two curves are plotted for two different values of the 
 decorrelation length: solid for $\delta_1=0.3$ years and broken for 
 $\delta_2=0.4$ years. }
 \label{Fig3} 
\end{figure}

Now we again apply the dispersion spectra method to the `clean' data set, i.e. a second
iteration is made. The 
result is surprising: $\Delta t_{B-A}=-0.38$ years. The technique should 
converge to a value near to that of the first result, if the previous estimation was correct and 
the technique is self-consistent. For consistency, we repeat this
analysis assuming a time delay of $-0.38$ years, i.e., a third iteration. 
The result is again unexpected: we recover the previous
value of $-0.73$ years. These results can be seen in Fig. \ref{Fig3},
upper panel (dispersion with all points), middle panel (borders and gap 
corrected around 1 year) and bottom panel (borders and gap corrected around 
half a year) where the minimum of the function gives the time delay. The 
solid and broken lines in each figure correspond to two slightly 
different decorrelation lengths ($\delta_1=0.3$ years, $\delta_2=0.4$ years).

This clearly means that the method is not self-consistent when applying it to 
the current data set. The dispersion spectra method is very sensitive to individual 
points, and in poorly sampled sets such as this one, these points are critical. 
It is obvious that we need better techniques for the determination of the time delay. 
But these techniques must not be interpolating ones because the lightcurves have 
lots of variability and wide gaps, and any simple interpolation scheme might introduce 
spurious signals.

\subsection{Techniques based on the discrete correlation function (DCF)}
\subsubsection{Reasons for `clean' datasets}
Many authors have applied different versions of the DCF since it was
introduced by Edelson \& Krolik (1988; hereafter EK88). Here we have
selected five of them. These techniques take into account the global behavior 
of the curves, rather than `critical points'. But in order to well apply all these
methods one has to eliminate border effects and gaps as described previously. If
one does not do this, one will lose signal in the peak of the DCF and secondary
peaks could appear, which can bias the final result. We will 
demonstrate this last point later 
(Fig. \ref{Fig7}, described in Sect. \ref{CEDCF}, is used for this 
purpose).

\subsubsection{Standard DCF plus a parabolic fit.}
\label{sdcf}
First we apply the usual form of the DCF to the data set. We briefly recall 
the expression, following EK88:
\begin{equation}
\label{ec-dcf}
DCF(\tau)=\frac{1}{M}\sum_{ij}\frac{(a_i-\bar{a})(b_j-\bar{b})}{\sqrt{
(\sigma^2_a-\epsilon^2_a)(\sigma^2_b-\epsilon^2_b)}},
\end{equation}
where $M$ is the number of data pairs ($a_j, b_j$) in the bin associated with the 
lag $\tau$, $\epsilon_x$ the measurement
error, $\sigma_x$ the standard deviation and $\bar{x}$ the mean of x. It gives
the cross correlation between both components at lag $\tau$ by considering bins 
that include all pairs of points ($a_j, b_j$) verifying  
$\tau-\alpha\leq(t_j-t_i)<\tau+\alpha$, where $\alpha$ is the bin semisize.
In DCF-based techniques, one always needs to find a compromise between 
the bin size and the error for each bin: increasing the former decreases the latter, 
but resolution with respect to $\tau$ is lost. The result of applying this 
procedure to the HE~1104$-$1805 data
is a function with a few points and without a prominent feature around the peak, 
because of our sparse sampling. The position of the peak gives the time 
delay: $\Delta t_{B-A}=-0.91$ years.

\begin{figure}[hbtp]
 \centering
 \epsfxsize=6 cm
 \rotatebox{-90}{\epsffile{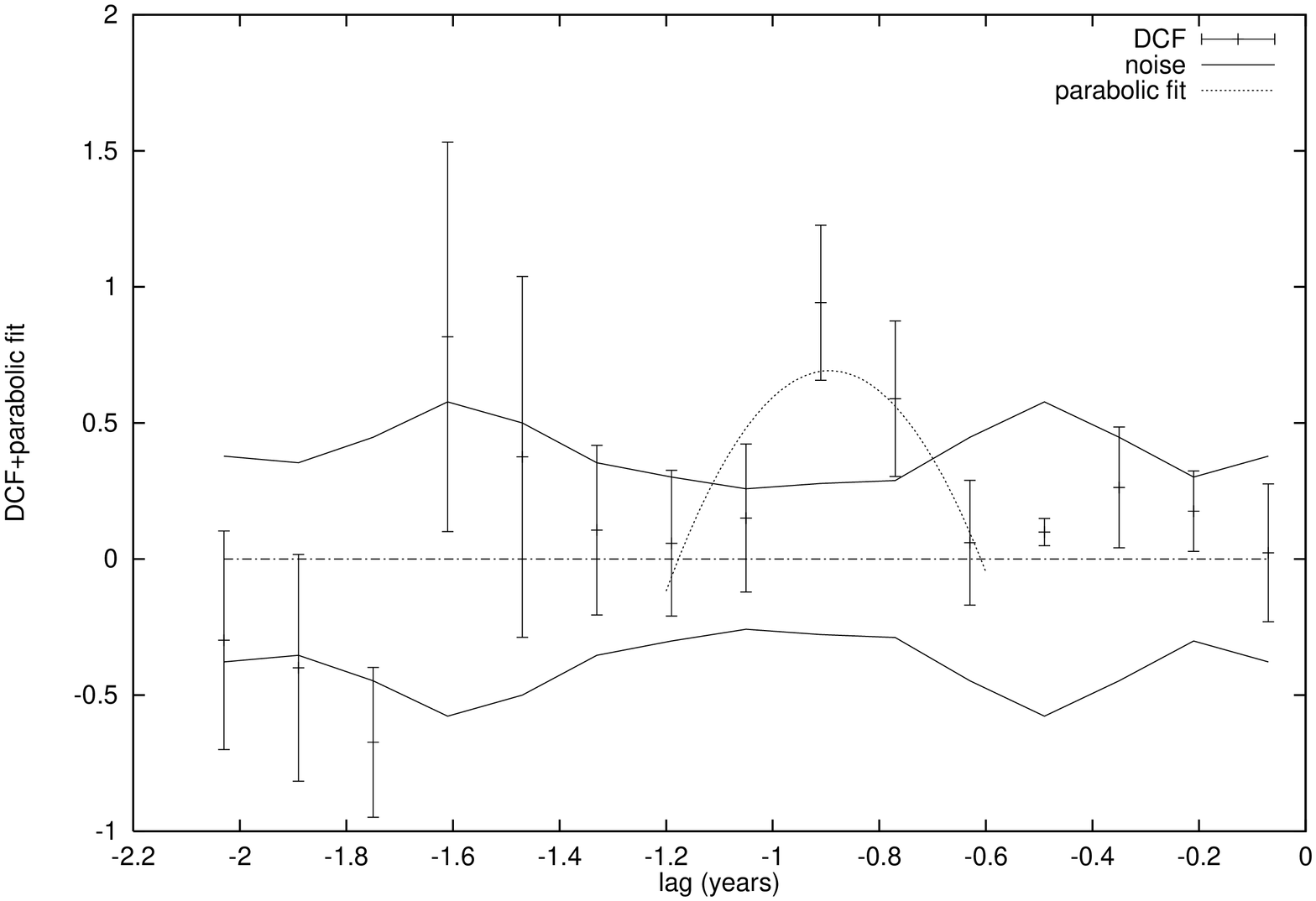}}
 \caption[]{The standard DCF and an added fit are shown in this figure. The peak
 is located at $-0.89$ years ($-0.91$ without fit) using a bin semisize 0.07
 years. The continuous lines are the noise levels and the zero level is also plotted. 
 Only two points defining the peak are outside the noise band.}
 \label{Fig4}
\end{figure}

A modification of this method was suggested by Leh\'ar et al. (1992). 
A parabolic fit to the peak of the function was proposed to solve the
problem of not resolving the peak. Doing this fit, we obtain a time 
delay of $\Delta t_{B-A}=-0.89$ years. These results are shown in Fig. \ref{Fig4}. 
The noise level is computed as $\sqrt{M}$, $M$ being the number of
pairs in each bin. The problem in this case is that the peak of the function is
defined with only two points above the noise level. We used a bin semisize 
of $\alpha=0.07$ years. Increasing the bin semisize to $\alpha=0.14$ years the result is not 
better in the sense that the peak is defined by only one point, and the fit does 
not modify the location of this peak. The obtained value for the time delay in 
this case ($\alpha=0.14$ years) is $\Delta t_{B-A}=-0.84$ years.

\subsubsection{Locally normalized discrete correlation function (LNDCF): 
averaging in each bin}
\begin{figure}[hbtp]
 \centering
 \epsfxsize=6 cm
 \rotatebox{-90}{\epsffile{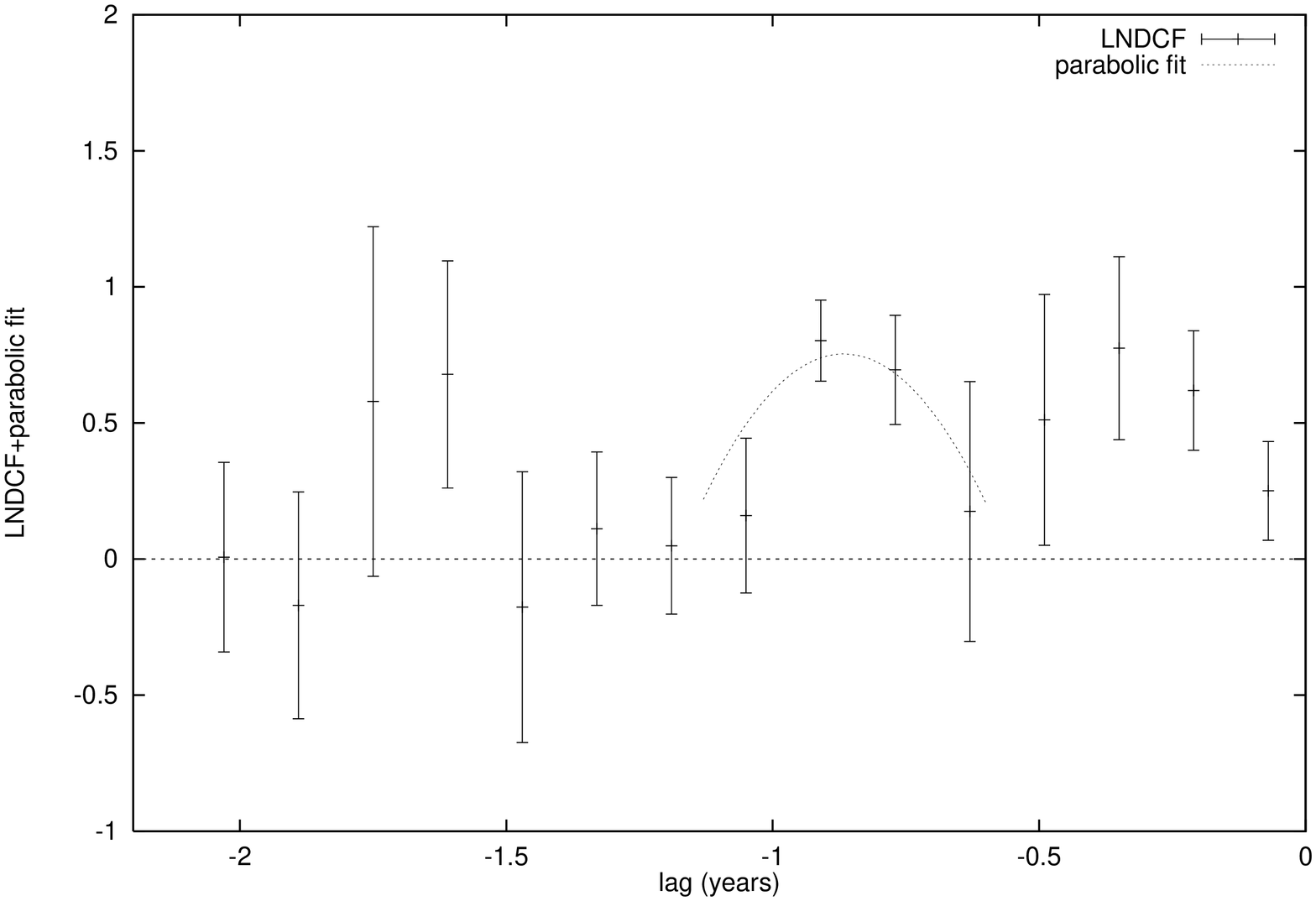}}
 \caption[]{The LNDCF is evaluated with a 0.07 years bin 
 semisize and the peak is fitted with a parabolic law. The result is a time delay 
 $\Delta t_{B-A}=-0.87$ years ($-0.91$ years without the fit). A secondary 
 peak appears at $-0.35$ years, although with larger error bars. This peak was the feature
 that "confused" the dispersion spectra.}
 \label{Fig5}
\end{figure}

The locally normalized discrete correlation function was also proposed by Leh\'ar 
et al. (1992). Its main difference to the simple DCF is that it computes the means and
variances locally (i.e. in each bin):
\begin{equation}
\label{ec_lndcf}
LNDCF(\tau)=\frac{1}{M}\sum_{ij}\frac{(a_i-\bar{a}_\ast)(b_j-\bar{b}_\ast)}
{[(\sigma^2_{a_\ast}-\epsilon^2_a)(\sigma^2_{b_\ast}-\epsilon^2_b)]^{1/2)}},
\end{equation}
computing the sum over all pairs where $\tau-\alpha\leq(t_j-t_i)<\tau+\alpha$.
The mean, $\bar{x}_\ast$, and the standard deviation, $\sigma^2_{a_\ast}$, are 
calculated for each bin. Again a parabolic fit is needed 
for a more accurate value of the peak, which then gives the time delay. For the same
reasons as in Sect. \ref{sdcf} we choose a bin semisize $\alpha=0.07$ years. 
The result is shown in Fig. \ref{Fig5}. As in the case of the standard DCF, the
peak is just defined by two points. The obtained time delay in this case is 
$\Delta t_{B-A}=-0.87$ years (the value without the fit is $-0.91$ years).
Furthermore, a secondary competing peak appears at $-0.35$ years, with more 
points, although these points have larger errorbars. This is an interesting
aspect, because 
it was this secondary peak which `confused' the dispersion spectra technique and it 
may suggest a close relation between these two techniques (both favour
`local' behaviour of the signals, rather than `global' ones). We will investigate 
this possible relation in a future, more general paper.

In any case, the poorly defined peak means the technique is again quite sensitive to our 
poor sampling. We look for a method less sensitive to this problem. The two following 
techniques are two different ways of trying to solve the problem of not having many 
points around the prominent peak.

\subsubsection{Continuously evaluated discrete correlation function (CEDCF): 
overlapping bins in the DCF}
\label{CEDCF}
The continuously evaluated bins discrete correlation function was introduced by 
Goicoechea et al. (1998a). The difference to the standard way of computing the 
DCF in this method is that the bins are non disjoint (i.e. each bin ovelaps 
with other adjacent bins, see  paragraph \ref{sdcf} where the bins do not 
overlap each other). One has to fix the distance between the centers 
of the bins in addition to their width. In this way it is possible to evaluate 
the DCF in more points, having a more continuously distributed curve. We will 
have also more significant points around the peak, i.e. above the noise level, 
and there is no need for fitting.
\begin{figure}[hbtp]
 \centering
 \epsfxsize=6 cm
 \rotatebox{-90}{\epsffile{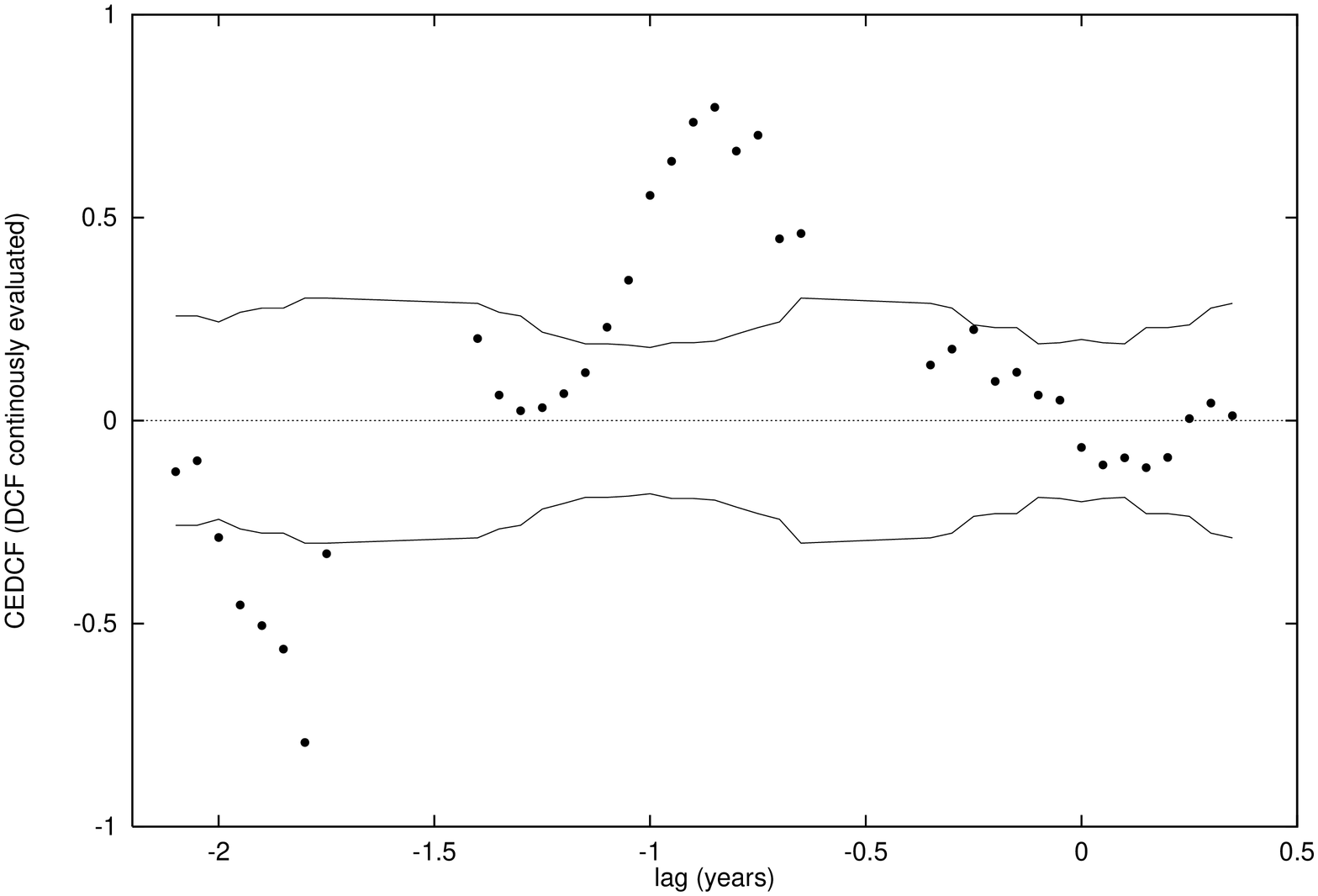}}
 \epsfxsize=6 cm
 \rotatebox{-90}{\epsffile{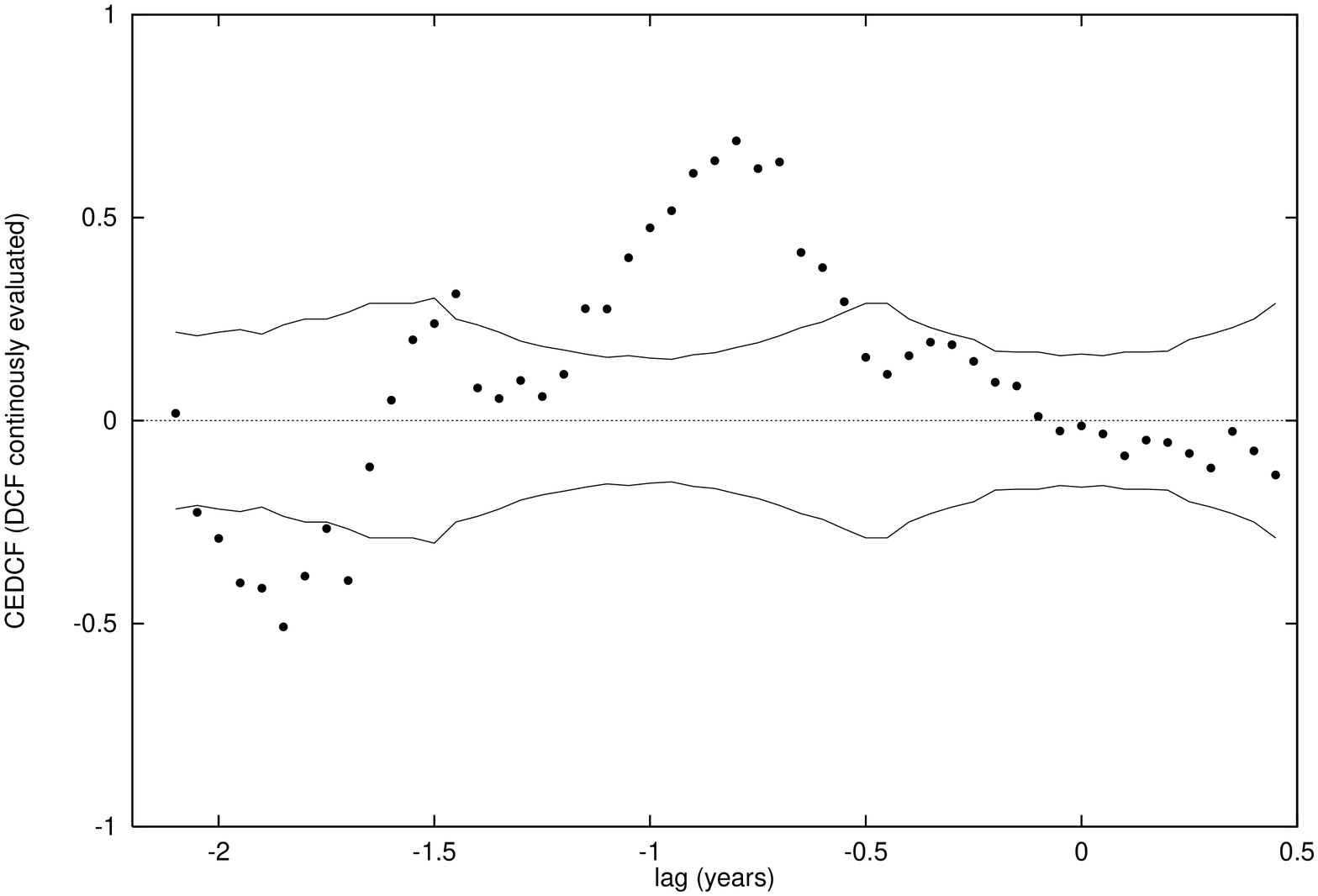}}  
 \caption[]{The CEDCF is a DCF more continuously evaluated. \emph{Top panel}: 
 using a bin semize of $\alpha=0.14$ years we obtain a peak at $-0.85$ years with a good 
 signal-to-noise ratio equal to 3.9. \emph{Bottom panel}: with a bin semisize equal 
 to $\alpha=0.21$ years, the peak is at $-0.80$ years. Although it seems that the 
 function is better defined, i.e. with more points, the signal-to-noise ratio 
 at the peak is 3.8. The continuous lines are the noise levels in both panels
 (cf.\ also Fig.~\ref{Fig7}).}
 \label{Fig6}
\end{figure}
Selecting the distance between the centers of the bins is again a matter of 
compromise: increasing the distance means needing wider bins and, thus, losing
resolution. The adopted time resolution should depend on the sampling; it seems
reasonable to select a value slightly less then the inverse of the highest frequency 
of sampling ($1/f\simeq0.1$ years). We choose 0.05 years as the best value for the
distance between bin centers and two values for bin semisizes: $\alpha=0.14$ 
and $\alpha=0.21$
years. The overlapping between bins allows us to consider slightly wider bin sizes.
We plot the results in Fig. \ref{Fig6}, upper and lower
panel, respectively. The continuous lines are the noise levels. 
The $\alpha=0.14$ years semisize shows a peak at $-0.85$ years, whereas with
the $\alpha=0.21$ years semisize the peak is at $-0.80$ years.

Now we need a good reason for preferring one over the other bin size. This
reason could be the signal-to-noise ratio of the peak: in the first case 
$\alpha=0.14$ years, $S/N=3.9$, and in the second $\alpha=0.21$ yrs, $S/N=3.8$. 
Clearly, the difference of these two values is not high enough to conclude that one
of them is the best.

\begin{figure}[hbtp]
 \centering
 \epsfxsize=6 cm
 \rotatebox{-90}{\epsffile{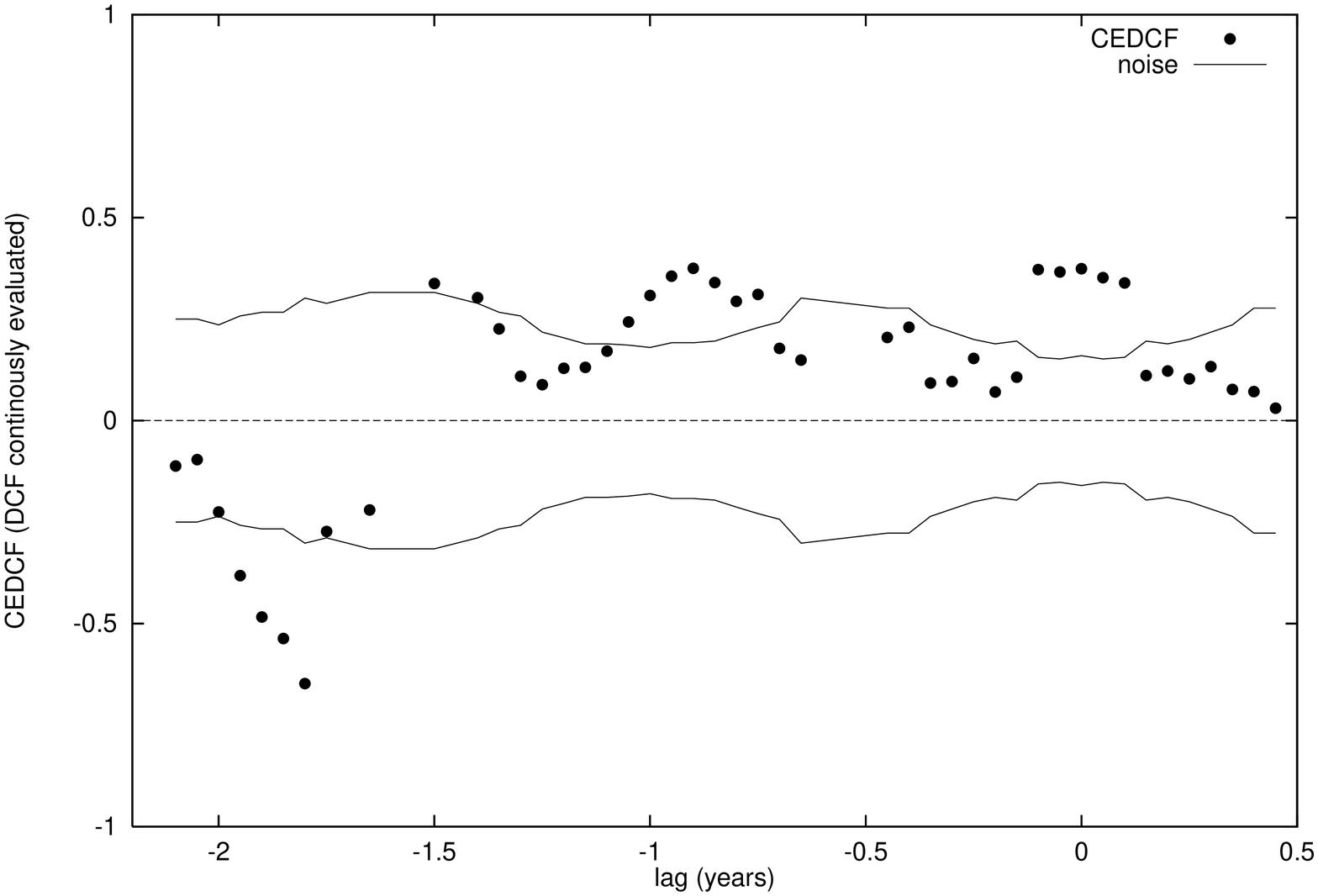}}
 \caption[]{Not eliminating borders can be crucial in DCF-based methods.
 Here the CEDCF has been computed with the original data set, i.e. using all points.
 There is a peak at $-0.90$ years, with a signal-to-noise value of 1.95. Other
 points around a secondary peak located at time zero describe another feature. 
 The great amount of information lost in the main peak is obvious.}
 \label{Fig7}
\end{figure}

In spite of the insignificant difference in this case, we notice that the 
signal-to-noise ratio is an important aspect and it 
is here where we justify the need for using `clean' data sets, i.e. 
border effects and gaps corrected. In Fig. \ref{Fig7} we plot the CEDCF 
for the original dataset (without any correction): the peak is located at 
$-$0.90 years, but the signal-to-noise is 1.95!. The main peak loses signal 
recovered by a secondary competing peak around lag zero and by the wings. 
Although this secondary peak is very unlikely to be the delay peak, Fig. 
\ref{Fig7} cannot solve this ambiguity, which demonstrate that border effects can be 
dramatic in some cases. In Sect. \ref{delta2} we will discuss the criteria to 
select a particular bin size.
\begin{figure}[hbtp]
 \centering
 \epsfxsize=5.8 cm
 \rotatebox{-90}{\epsffile{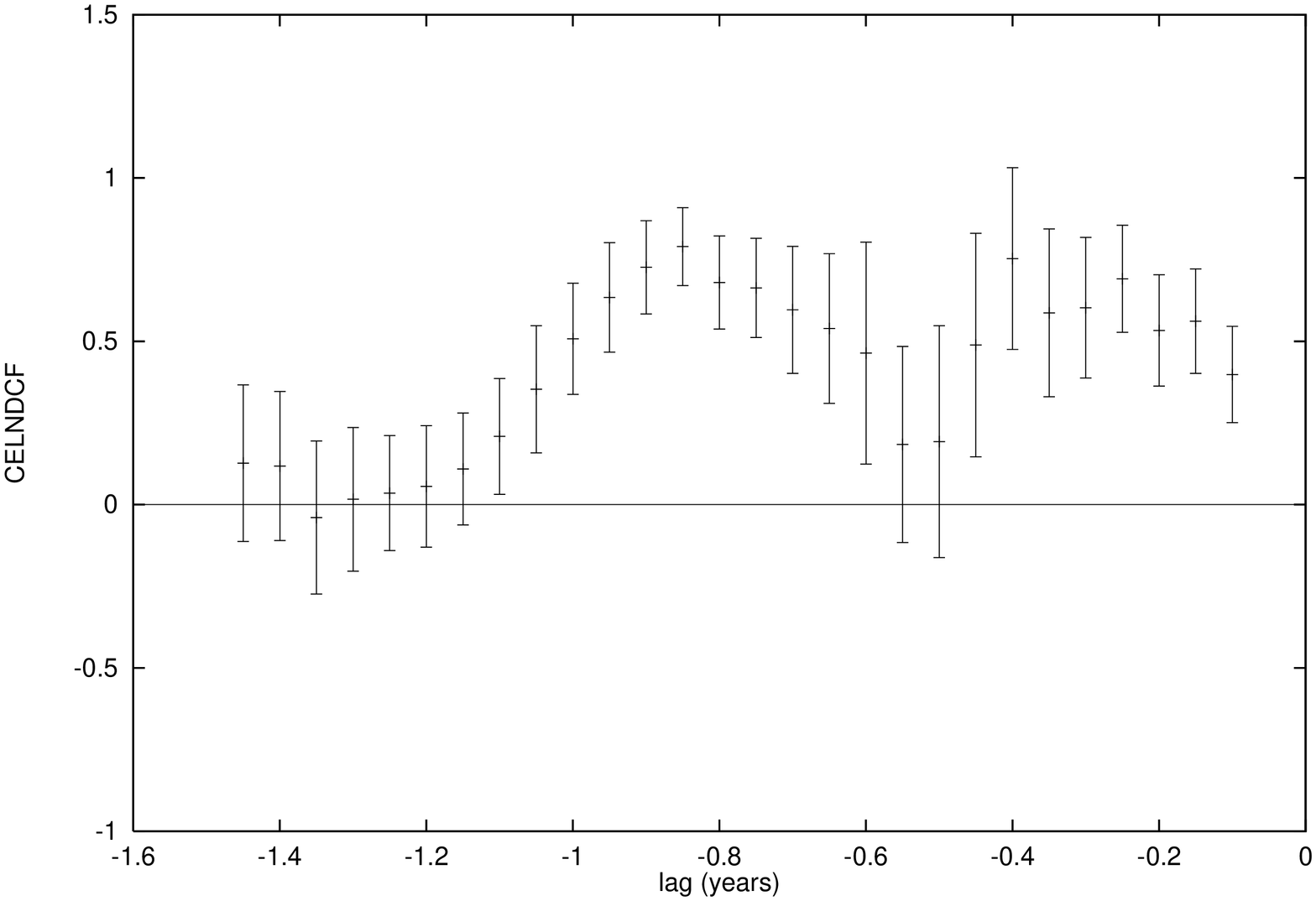}}
 \epsfxsize=5.8 cm  
 \rotatebox{-90}{\epsffile{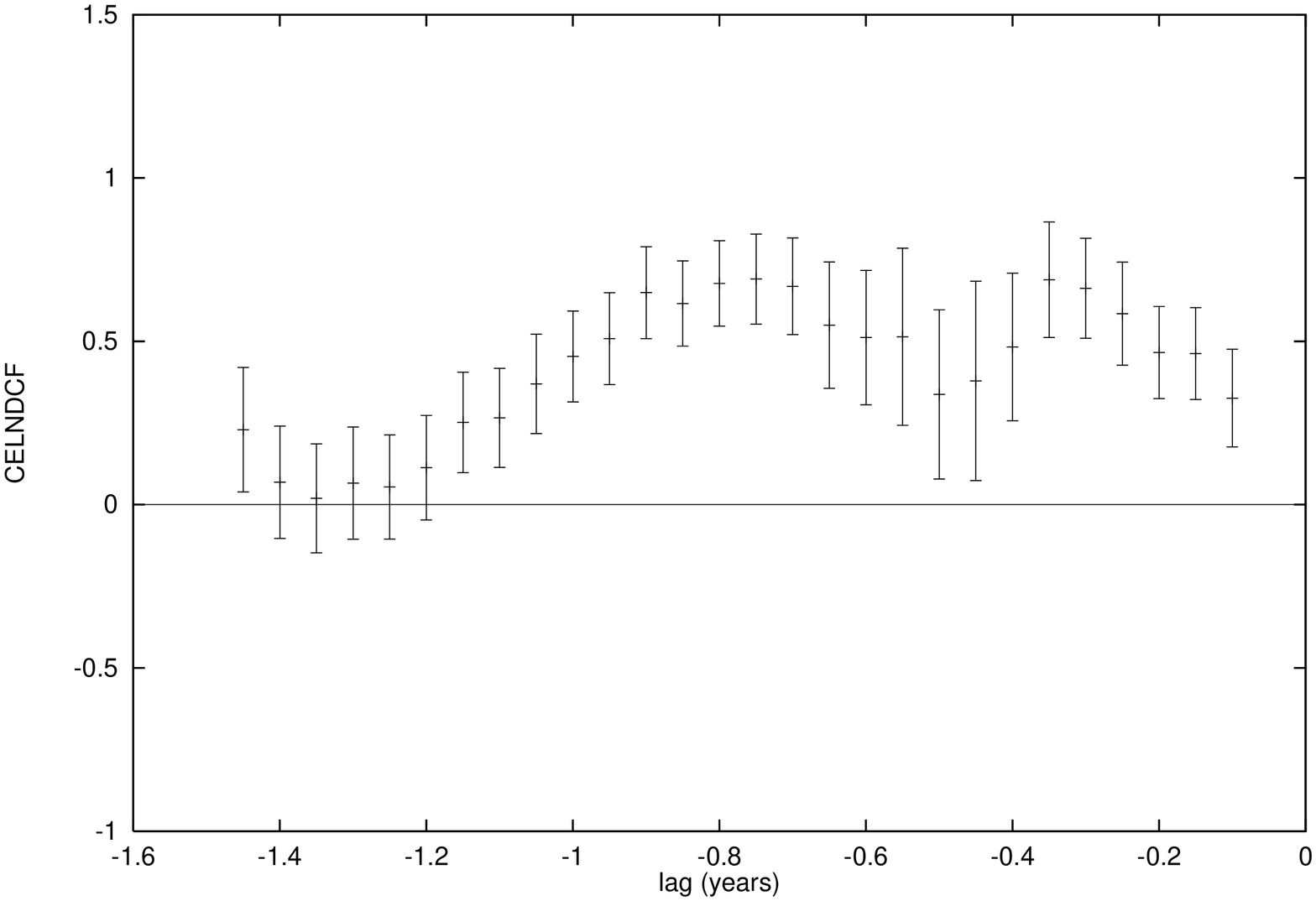}}
 \caption[]{\emph{Top panel}: The CELNDCF is evaluated with $\alpha=0.14$ years 
 bin semisize and a distance between bin centers of 0.05 years. The result is 
 a time delay $\Delta t_{B-A}=-0.85$ years. \emph{Bottom panel}: The CELNDCF 
 computed with $\alpha=0.21$ years bin semisize. The distances between the bin 
 centers is also 0.05 years. The peak is obtained at $-0.75$ years where it is 
 assumed to be the time delay.}
 \label{Fig8}
\end{figure}
\begin{figure}[hbtp]
 \centering
 \epsfxsize=9 cm
 \rotatebox{0}{\epsffile{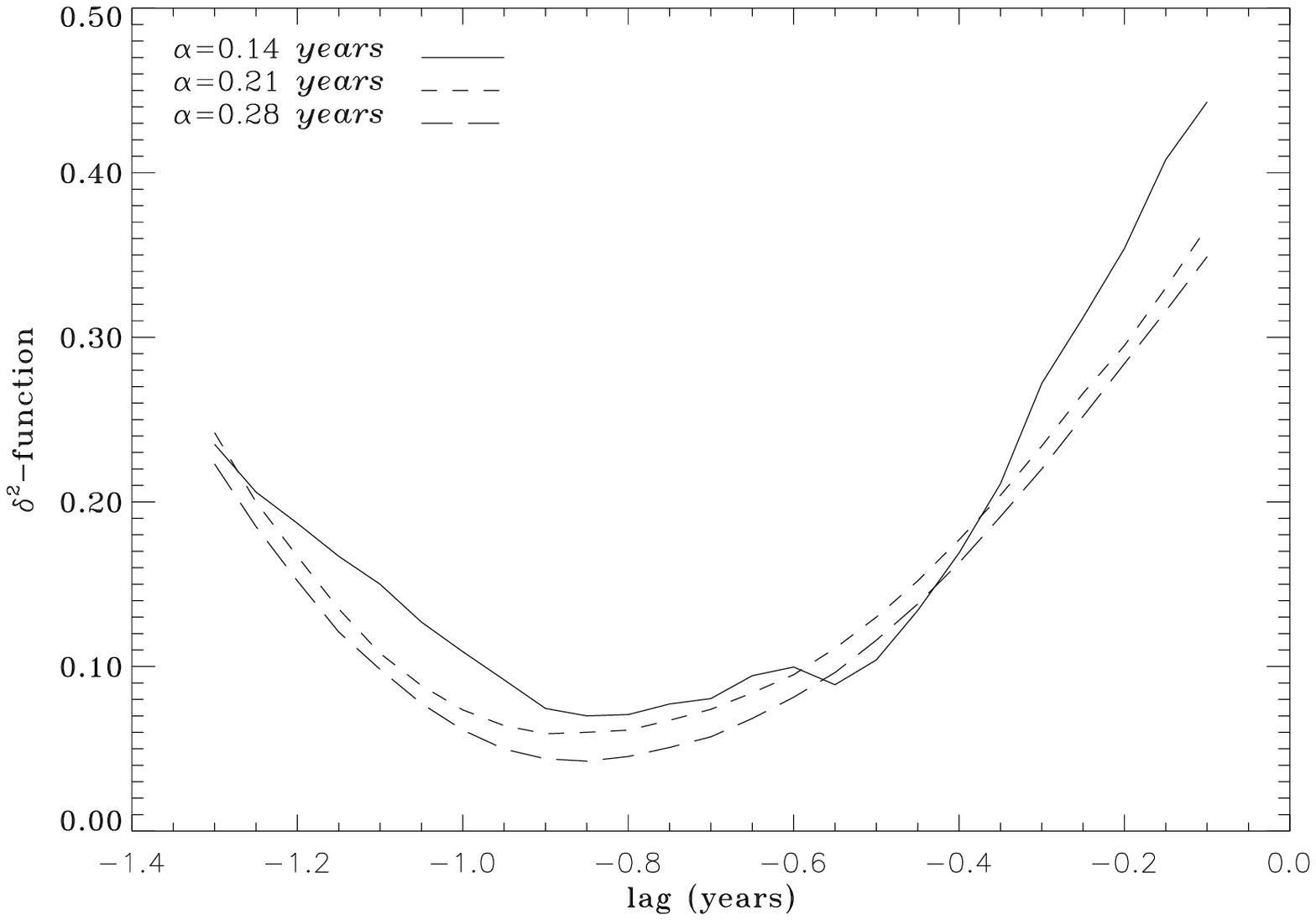}}
 \caption[]{The $\delta^2$ function for three
 different values of the bin semisize $\alpha$: solid line 0.14 years, short
 dashed 
 0.21 years and long dashed 0.28 years. Since $\delta^2_{\rm min}(\alpha=0.28)
 <\delta^2_{\rm min}(\alpha=0.21)<\delta^2_{\rm min}(\alpha=0.14)$, the features in $\delta^2$
 for $\alpha=0.14$ years is unlikely to be an artifact (see text for more details).} 
 \label{Fig9}
\end{figure}

\subsubsection{Continously evaluated bins and locally normalized discrete
 correlation function (CELNDCF): overlapping bins in the LNDCF}
\label{CELNDCF}
To our knowledge, this technique has not been applied before, but it seems a
natural step as a combination of the two former techniques (i.e., the LNDCF and the
CEDCF). From the one side, we use Eq. (\ref{ec_lndcf}) for computing the DCF,
i.e., it is a locally normalized discrete correlation function. From the other
side, we use the idea of overlapping bins described in Sect. \ref{CEDCF}. Thus,
the final result is a `continuously evaluated bins and locally normalized discrete
correlation function' (CELNDCF). Again, we fix the distances between the bins and
also their width. The result will be a function similar in shape to the LNDCF in Fig.
\ref{Fig5} but with more points evaluated.

The method was applied for three different values of the bin semisize: 0.07,
0.14 and 0.21 years. The first value is not a good choice, it gives
relatively large errorbars for the points of the CELNDCF, since the number of 
points per bin is low. 
Selecting the last two values, i.e. $\alpha=0.14$ yrs. and $\alpha=0.21$ yrs., 
we obtain Fig. \ref{Fig8}. The 
first one gives a time delay  of $\Delta t_{B-A}=-0.85$ years and the second 
one a value of $\Delta t_{B-A}=-0.75$ years. This second
result is very close to the first estimation in W98. The reader can easily
compare the results with and without overlapping bins (Fig. \ref{Fig8} and Fig.
\ref{Fig5}, respectively) and clearly see the advantages of this second 
procedure. Nevertheless, there is a relatively large difference between 
selecting one or the other value of the bin semisize (i.e. $\alpha=0.14$ years 
vs. $\alpha=0.21$ years). This means the technique is also very sensitive to 
the poor sampling. The next and final technique will clarify which is the best 
bin size selection. 
\begin{figure}[hbtp]
 \centering
 \epsfxsize=5.75 cm
 \rotatebox{-90}{\epsffile{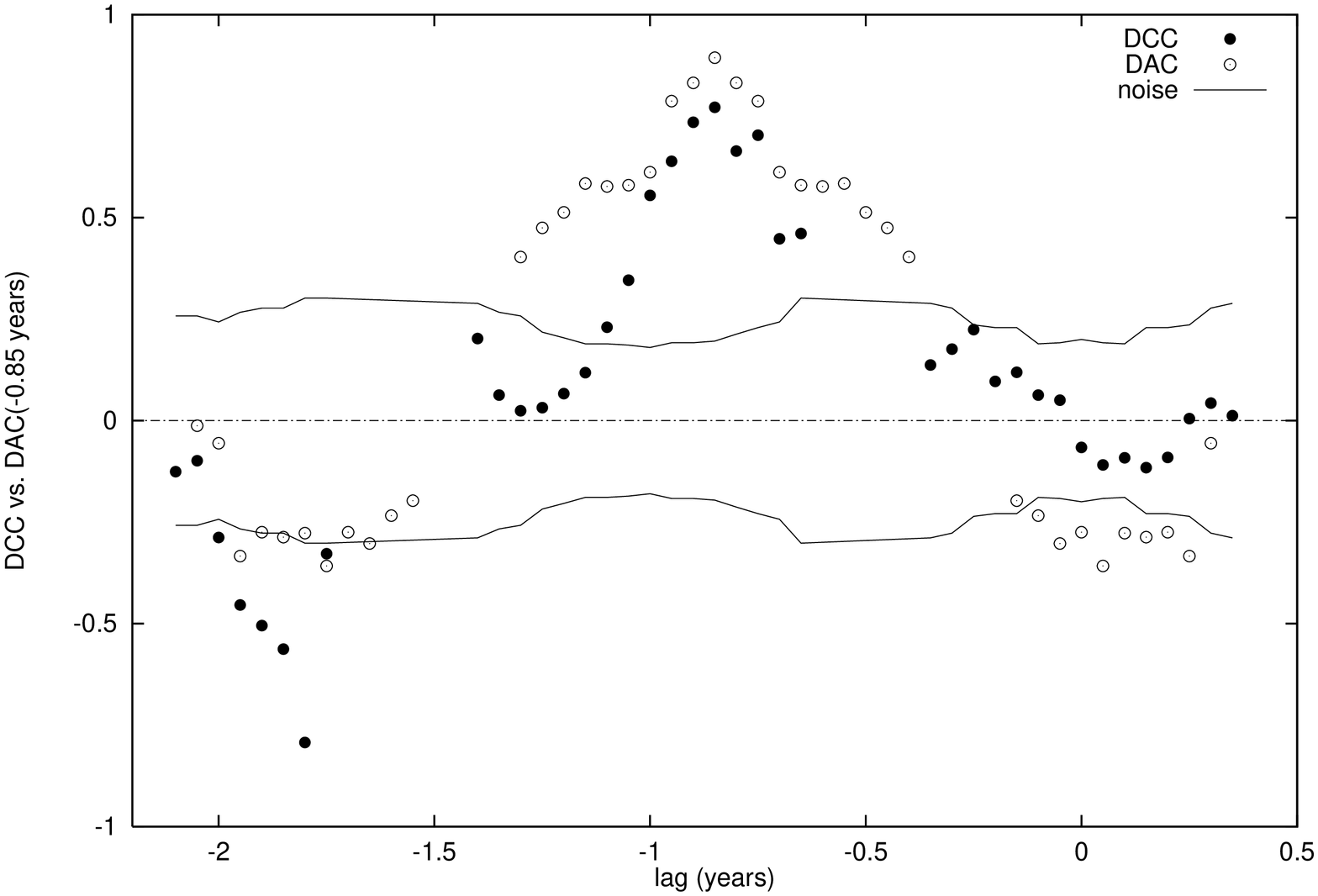}}
 \epsfxsize=5.75 cm
 \rotatebox{-90}{\epsffile{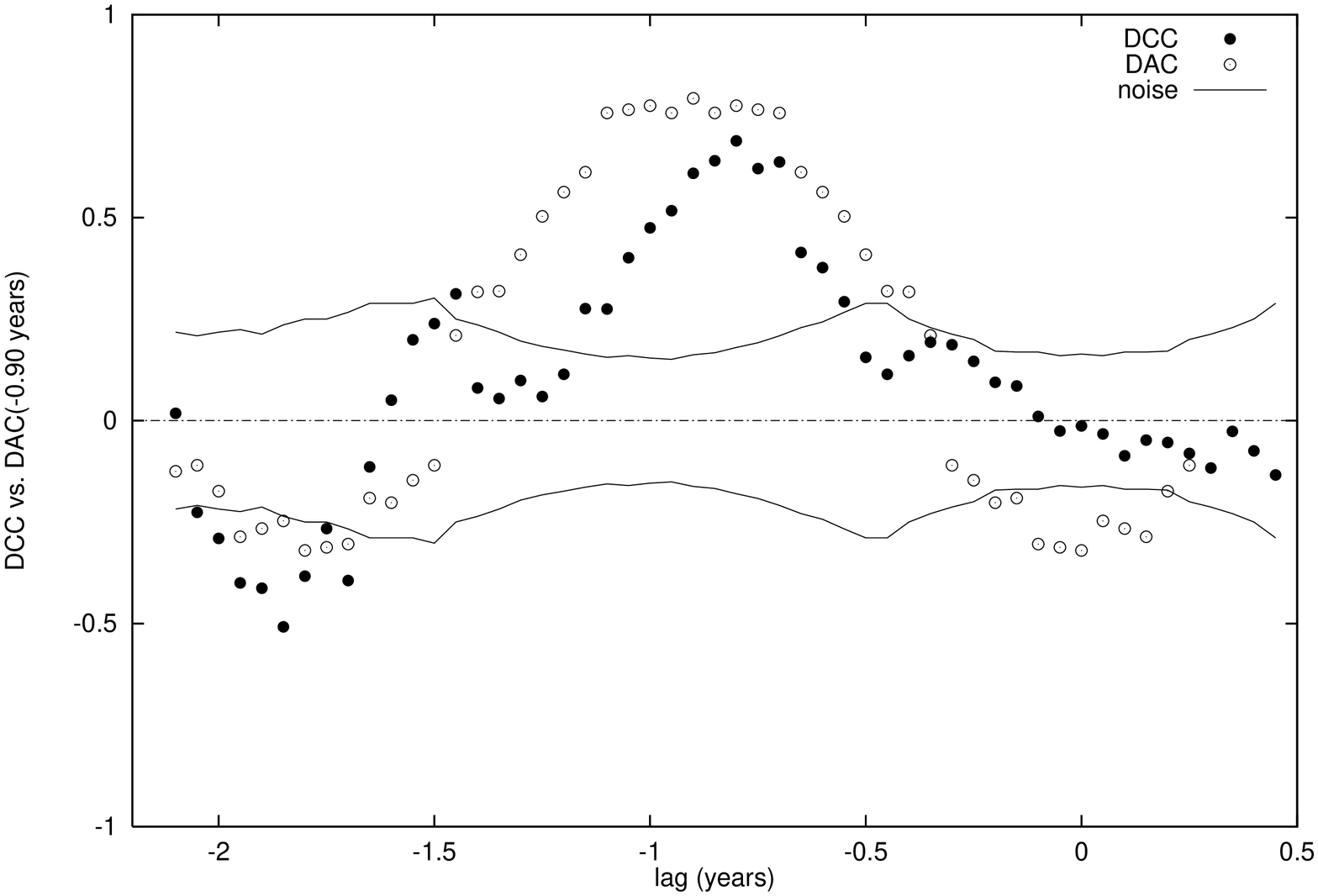}}
 \epsfxsize=5.75 cm
 \rotatebox{-90}{\epsffile{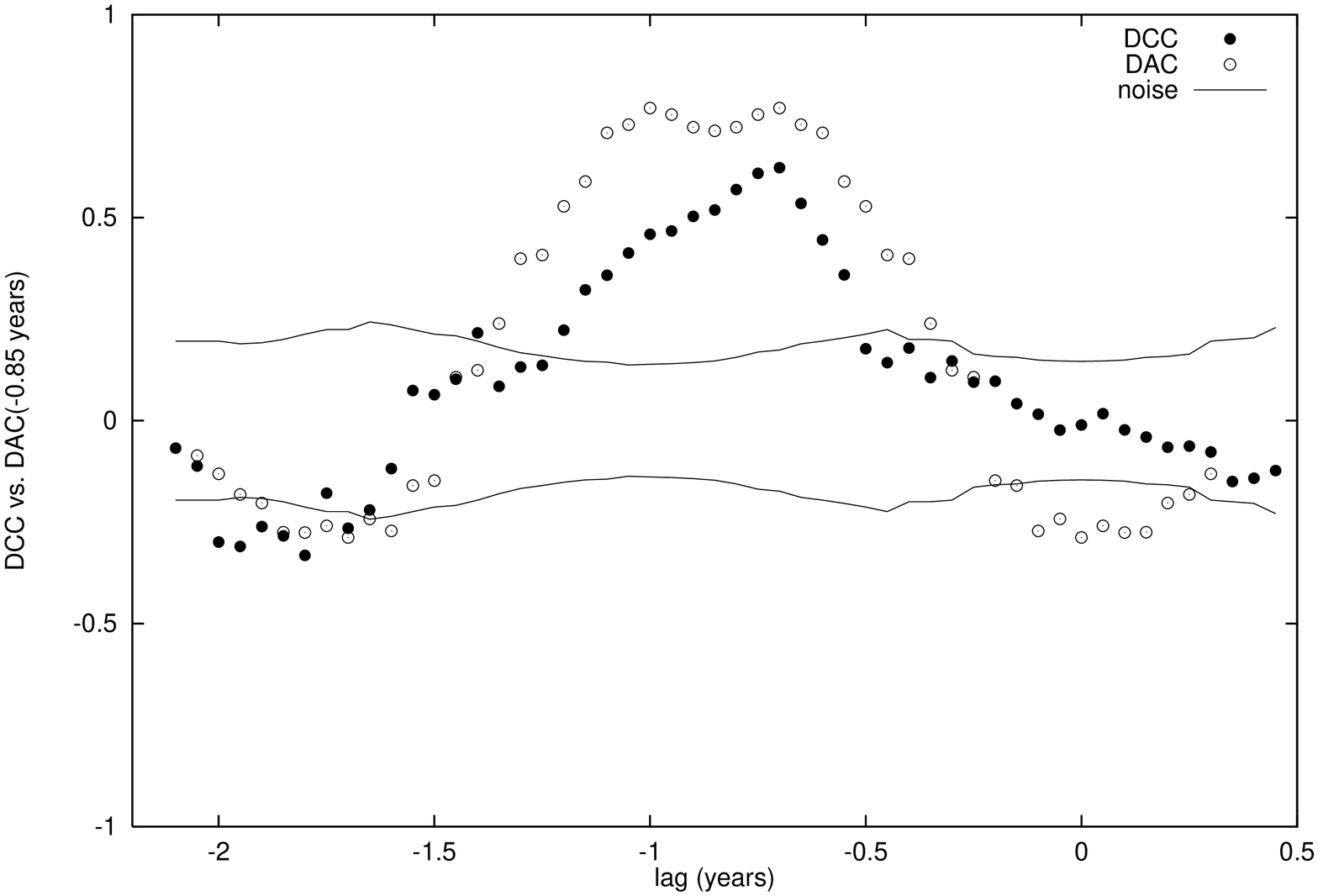}}   
 \caption[]{\emph{Upper panel}: both DCC (filled circles) and DAC (open circles) 
 are plotted. The bin semisize is $\alpha=0.14$ years and the DAC has been 
 shifted by $-0.85$ years, which is the value for the time delay obtained with the 
 $\delta^2$ technique. \emph{Middle panel}: the bin semisize is now 
 $\alpha=0.21$ years. DAC (open circles) has now been shifted by $-0.80$ years, 
 which is the value obtained with the $\delta^2$ technique. The bin semisize 
 is now $\alpha=0.21$ years. \emph{Bottom panel}: for $\alpha=0.28$,
 $\delta^2_{min}=-0.85$ again, so the DCC (filled circles) is shifted by that value.
 In the three subfigures the solid lines indicate the noise levels. The best
 agreement between DCC and DAC is for $\alpha=0.14$ years (upper pannel).}
 \label{Fig10} 
\end{figure}
\subsection{The $\delta^2$ technique: a comparison between the cross correlation 
function and the autocorrelation function}
\label{delta2}
The following method, called $\delta^2$, was introduced by \cite{Goico98b}
and \cite{Serra99}. Its expression is
\begin{equation}
\delta^2(\theta)=\frac{1}{N}\sum^N_{i=1}S_i[DCC(\tau_i)-DAC(\tau_i-\theta)]^2
\end{equation}
where $S_i=1$ if $DCC(\tau)$ and $DAC(\tau_i-\theta)$ are both defined and 
$S_i=0$ otherwise. The DCC is the continuously evaluated discrete
correlation function, and the DAC is the discrete autocorrelation function.
The method uses the DCC and the DAC of one of the
components, and tries to get the best match between them by minimizing its 
difference. If one has two equal signals, these functions must be identical. 
The $\delta^2$ function reaches its minimum $\theta_0=\Delta t_{B-A}$ at
the time delay. We note that the match of both functions is not a
match between their peaks, but rather a global match.

We have selected component B for computing the DAC, because component A has more 
variability (presumably due to microlensing). We compute $\delta^2$ for 
different values of the bin
semisize. Adopting a bin semisize $\alpha=0.14$, the function shows some
features and reaches its minimum at $-0.85$ years (see Fig. \ref{Fig9},
solid line). Now we compute $\delta^2$ for a bin semisize 
$\alpha=0.21$ years, which yields a minimum at $-0.90$ years (Fig. \ref{Fig9}, 
long dashed line). The question now is: are we loosing resolution using this last 
bin semisize ($\alpha=0.21$ years) or is this minimum at $-0.90$ years a better
estimate?
The reader could argue that 
$\delta^2_{\rm min}(\alpha=0.21)<\delta^2_{\rm min}(\alpha=0.14)$, 
so that the agreement between DAC and DCC is better for $\alpha=0.21$. This is
not so. Consider a bin semisize 
$\alpha=0.28$ years (Fig. \ref{Fig9}, short dashed line): We obtain a
minimum at $-$0.85 years while again 
$\delta^2_{\rm min}(\alpha=0.28)<\delta^2_{\rm min}(\alpha=0.21)$. This indicates that the
minimum located at $-0.85$ years with $\alpha=0.14$ years was not an artifact of
some noise features, but that these features are real. To clarify this, 
Fig. \ref{Fig10} shows the comparison between the DCC and DAC function for the
three different values of the bin semisize $\alpha$ (0.14, 0.21 and 0.28 years in
the upper panel, middle panel and bottom panel, respectively). Accordingly, we consider the
$\alpha=0.14$ years the best bin semisize and we analyse $\delta^2$ for
that value.

\begin{figure}[hbtp]
 \centering
 \epsfxsize=9 cm
 \rotatebox{0}{\epsffile{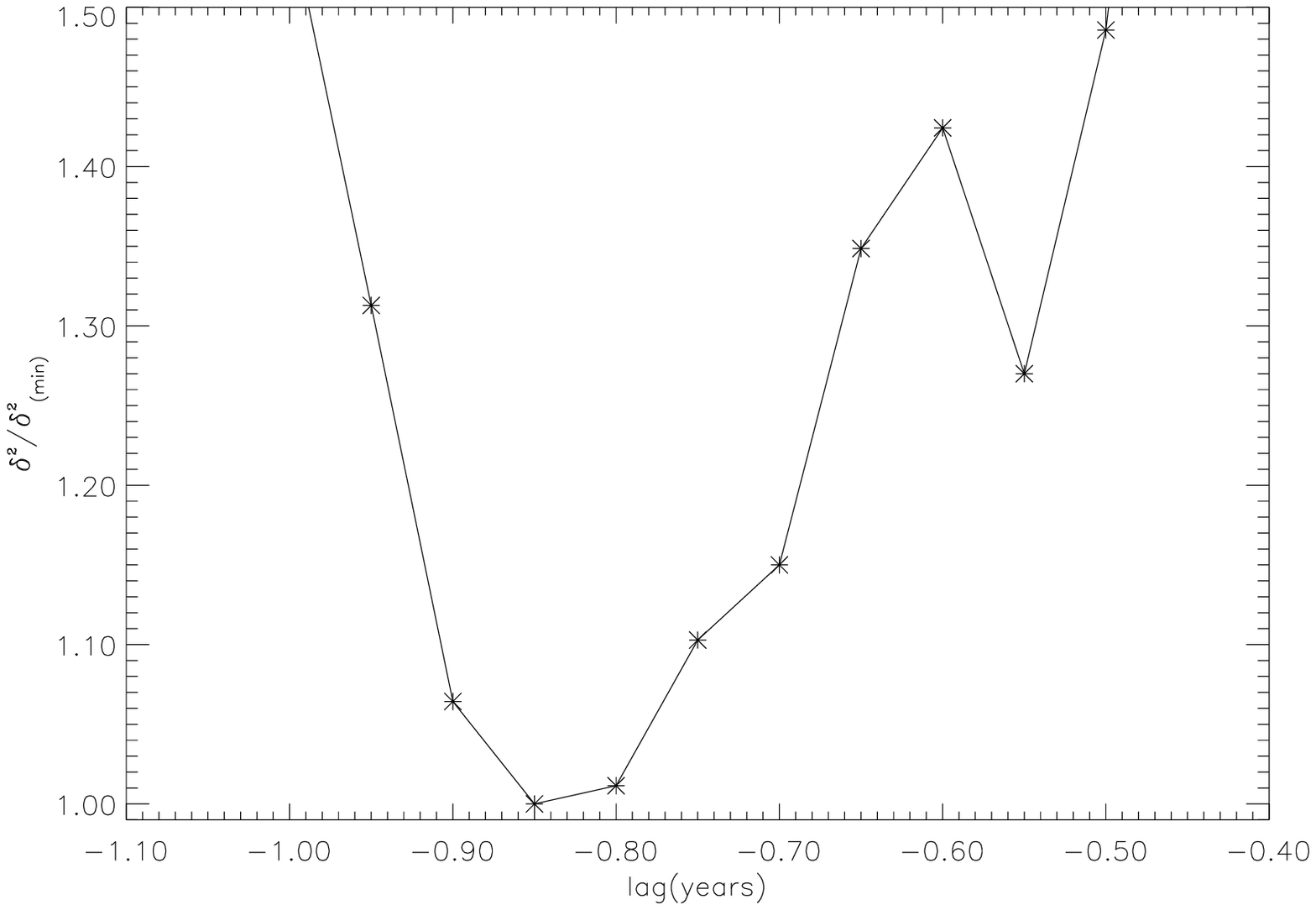}}
 \caption[]{The minimum of the $\delta^2$ function gives the time delay between
 the components. We have normalized it with its minimum. A secondary peak is present
 around $-0.55$, a value also considered by W98. The trend of the main feature is asymmetric,
 favoring values in the range [$-0.9$, $-0.7$], including several best estimates of the
 time delay from other techniques or binning.}
 \label{Fig11}
\end{figure}

In order to better study the features in the $\delta^2$ function, we plot it
normalized to its minimum in Fig. \ref{Fig11}. This figure is 
quite illustrative: (i) The minimum is reached at $-0.85$ years. (ii) The trend of the
main feature is asymmetric, with a relatively slow rise at the right hand side, favoring
values in the range $[-0.9, -0.7]$, including most of the estimates 
from other techniques or binning. (iii) A `secondary minimum' is present at
$-0.55$
years. This may be due to the fact remarked already by W98: for such a lag, the
observing periods of one component coincides with the seasonal gaps in the lightcurve
of the other. (iv) The feature in the range $[-0.3, -0.4]$ is not present,
meanning that this value is very unlikely (this was the value that appeared 
with dispersion spectra, LNDCF and CELNDCF methods).

\begin{figure}[hbtp]
 \centering
 \epsfxsize=6 cm
 \rotatebox{-90}{\epsffile{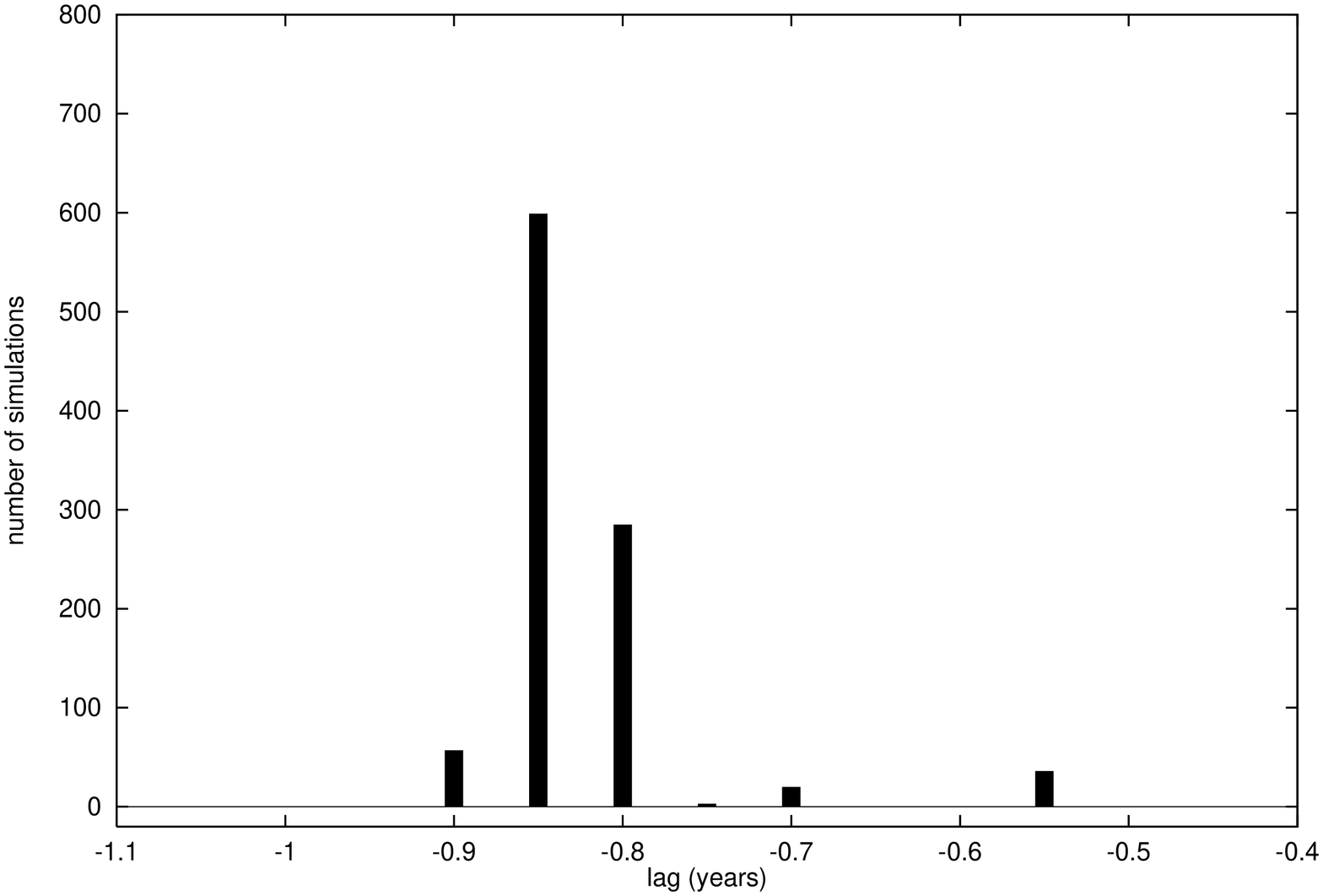}}
 \caption[]{Histogram of time delays obtained in 1000 Monte Carlo simulations, 
 using the $\delta^2$ technique. }
 \label{Fig12}
\end{figure}

\begin{figure}[hbtp]
 \centering
 \includegraphics[bbllx=60,bblly=99,bburx=360,bbury=380,width=8.5cm,
                  clip=true]{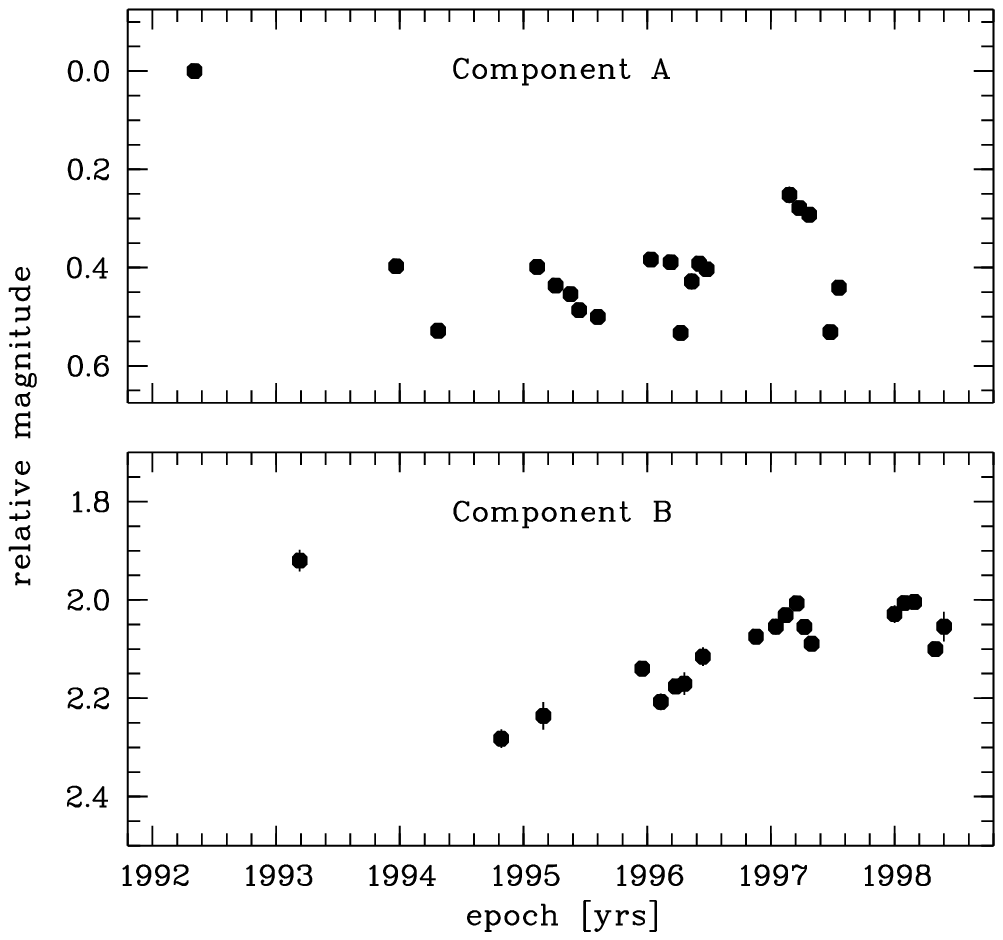}
 \caption[]{The original dataset with the component A shifted by the new 
 adopted time delay, $\Delta t_{B-A}=-0.85$ years.}
 \label{Fig13}
\end{figure}

To obtain an estimate for the formal error of this method, we used 1000 
Monte Carlo simulations. For each simulation we did the following: for each  
epoch $t_i$ we associated a value in magnitudes $x_{i}+\Delta x_{i}$, where 
$x_{i}$ is the observed value and $\Delta x_{i}$ is a Gaussian random variable with
zero mean and variance equal to the estimated measurement error. The histogram 
is presented in Fig. \ref{Fig12}. As it can be seen, the simulations 
reproduce all the information contained in the $\delta^2$ function in Fig. 
\ref{Fig11}: the most probable value is $-0.85$ years (599 simulations); it also 
appears in a number of simulations around $-0.90$ years (57 simulations),
$-0.80$ years (285 simulations), $-0.75$ years (5 simulations) and around 
$-0.70$ years (20 simulations). A few 
simulations (36) are also located around $-0.55$ years, which is very close to 
the one considered in W98 as spurious (a 
value around half a year). 
In any case, the simulations are in very good 
agreement with the information contained 
in the $\delta^2$ function. As 95\% of 
the simulations claim a time delay in the interval $[-0.90, -0.80]$, we can 
adopt a value of $\Delta t_{B-A}=-0.85\pm 0.05$ for the time delay of 
this system, with a 2$\sigma$ confidence level (formal or internal error). 
Fig. \ref{Fig13} shows the lightcurves with component A shifted the adopted 
time delay.

\section{Discussion}
\subsection{Comparison of the different techniques}
From the tour through the different techniques we can learn several
useful things. First of all, when only one technique is selected for 
deriving a time delay between two signals, it is important to check the internal
consistency of the method and its behaviour with a given data set. We have
demonstrated in Sect. \ref{borders} that dispersion spectra does not pass
this test at least in this case (see Fig. \ref{Fig3}). We have then applied and
discussed the 
discrete correlation function and several of its modifications. 
The standard DCF (Fig. \ref{Fig4}) 
had problems to properly define the peak in the case of very poorly sampled lightcurves; 
although a fit was proposed to solve this problem, there were only two points 
above the noise level in the best case and the fit was not very plausible. The 
LNDCF (Fig. \ref{Fig5}), based on locally normalized bins, 
had a similar behaviour and although the error bars of each point are smaller, 
the peak is not well defined either. The CEDCF (Fig. \ref{Fig6}) and the 
CELNDCF (Fig. \ref{Fig8}) worked better under these circunstances, but we found 
the problem of selecting the bin size; in the case of the CEDCF the difference 
between the two selected bin sizes was smaller than in the case of the CELNDCF. 
Finally applying the $\delta^2$ technique, we found a good reason for selecting 
one bin size: 
the match between the DAC and the DCC. The resulting estimate and its 
uncertainty include, as a `byproduct', the 
results of the rest of the techniques for the same bin size (except the dispersion 
spectra method which was not self-consistent). This fact is not the same as computing 
all the techniques and doing some statistics to obtain an uncertainty. This
frequently appears in the time delay determination literature, although it is 
not at all clear which was the weight of each technique when computing the 
final result. We note that for consistency we should apply a correction to the
original data set with the final adopted time delay of $-0.85$ years. Due to the
(very) sparse sampling of our data set, this
correction gives a reduced data set identical to the previous `clean' data set 
obtained with a correction of $-0.73$ years, so we do not need to repeat the 
whole process. The procedure is self-consistent.

It is important to notice that we have not meant to establish any general hierarchy
between all these techniques. The hierarchy is valid 
in our particular case study. Nevertheless, the idea, not new, of correcting border 
effects in the signals with first estimations has been proved to be a good 
procedure in DCF based techniques.

\subsection{Investigation of secondary minima/maxima}
In some of the techniques we have discussed and applied 
here for the data of HE1104~$-$1805, there appear 
secondary peaks/dips located at different values for the
time lags (see Fig. \ref{Fig5}, Fig. \ref{Fig8} and 
Fig. \ref{Fig11}). 
Here we investigate two obvious effects that might 
cause such behaviour, namely microlensing and sampling.
We do this only as a case study for the 
$\delta^2$ technique, but assume that our conclusions can
be generalized to the other methods as well.

\subsubsection{Microlensing}
Microlensing affects the two quasar lightcurves differently. 
That means that the two lightcurves will not be identical copies
of each other (modulo offsets in magnitude and time), but 
there can be minor or major deviations between them. 
On the other hand, experience from other multiple quasar systems
tells us that microlensing cannot dominate the variability, 
because otherwise there would be no way to determine a time delay
at all.
In any case, microlensing is a possible source of `noise' with
respect to the determination of the time delay.

A complete analysis of microlensing on this system is beyond the scope
of this article, and will be addressed in a forthcoming paper.
Here we present a simple, but illustrative,
approach to the way microlensing can effect the determination
of the time delay, and in particular its effect on the $\delta^2$ technique.
An `extreme' view of microlensing was investigated by 
Falco et al. (1991), who showed for the Q0957$+$561 system that it is
very unlikely that 
microlensing can mimic `parallel' intrinsic fluctuations 
causing completely wrong values for the time delay
correlations.  
But strong microlensing clearly affects 
the features of the cross-correlation function
(Goicoechea et al. 1998a). Depending on the
exact amplitude and shape of the microlensing
event, the main and secondary peaks of this
function can be distorted, 
possibly inducing wrong interpretations. 

\begin{figure}[hbtp]
 \centering
 \epsfxsize=9 cm
 \rotatebox{0}{\epsffile{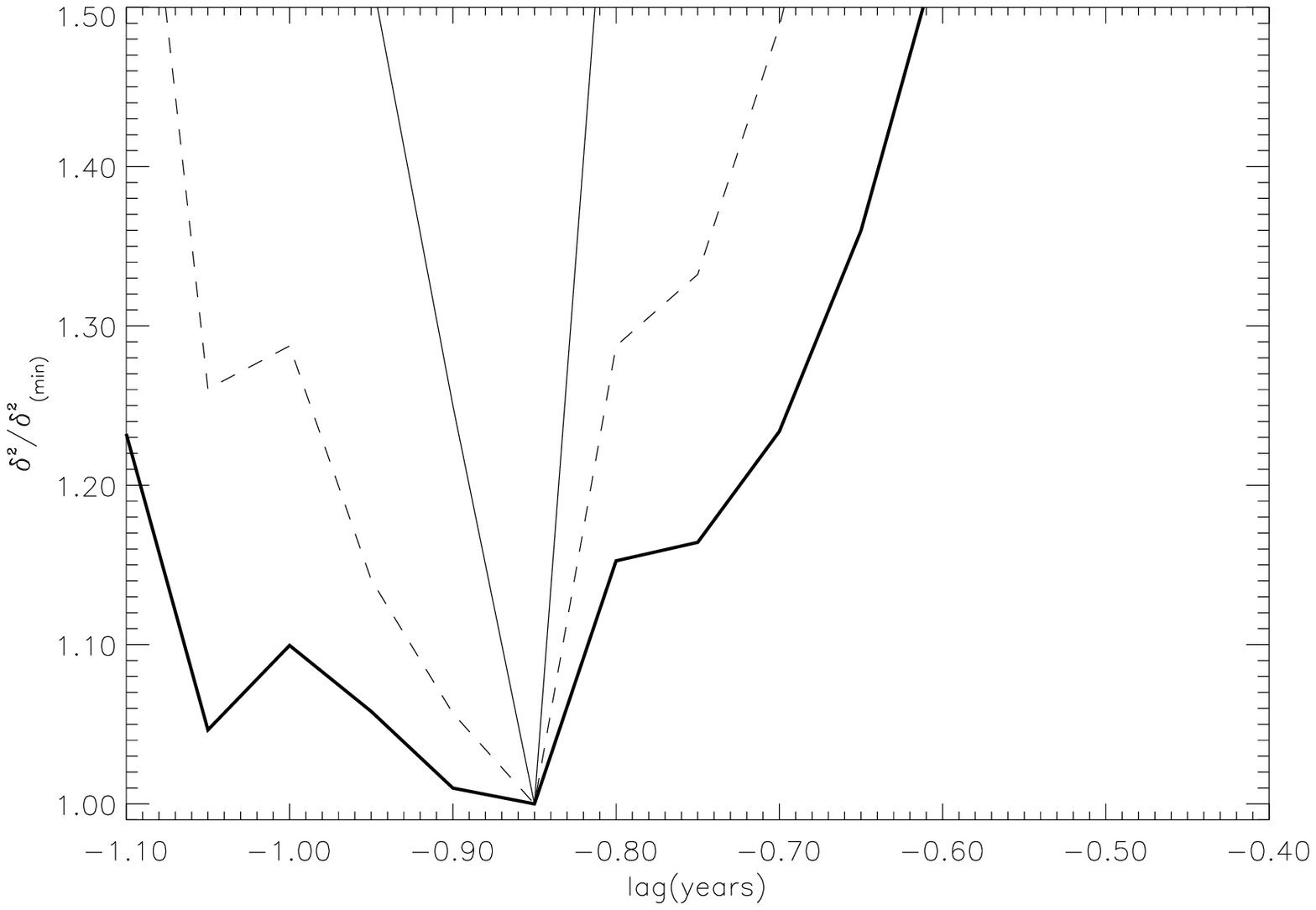}}
 \caption[]{We calculate the time delay between the lightcurves B and B$'$ with
 the $\delta^2$ technique. B$'$ is a copy of B, shifted $0.85$ years and 
 with a gaussian random process added.
 \emph{Thin solid line}: the gaussian random process has a standard deviation of
 $0.05$ mag. There are no secondary peaks. \emph{Dashed line}: If the standard
 deviation of the gaussian random process is $0.075$ mag., some secondary
 features appear. \emph{Thick solid line}: the $\delta^2$ normalized function is
 much more distorted, but the technique can calculate the shifted value of 
 $0.85$ years.}
 \label{Fig14}
\end{figure}

In order to study this effect here, we do the following: 
we consider the lightcurve of component B 
(assumed to reflect only intrinsic quasar variability)
and a copy of it, shifted by $0.85$ years, which we shall call B$'$. 
Obviously, any technique will give a time delay value 
of $\Delta t_{B-B'}=-0.85$ years between B and B$'$.
In the case of the $\delta^2$ technique, a very sharp minimum is located 
at this lag. 
Now we introduce artificial `microlensing' as a kind of 
Gaussian random process with zero mean and a certain standard deviation 
$\sigma_{\rm ML}$ to the lightcurve B$'$. 
We consider three cases: 
$\sigma_{\rm ML} =  0.050$ mag, $0.075$ mag and $0.100$ mag. 
Although microlensing is in general obviously not a random process (it depends
a bit on the sampling),
we use this simple approximation in order to study whether and 
how secondary peaks can appear in time delay determinations.
The resulting $\delta^2$-functions  can be seen in
Fig. \ref{Fig14}, which can be compared to Fig. \ref{Fig11}. 
It is very obvious that for the `smallest'  microlensing contribution
($\sigma_{\rm ML} =  0.050$ mag, thin solid line)
the minimum of the $\delta^2$ normalized function is still 
a very sharp feature. 
For the next case ($\sigma_{\rm ML} =  0.075$ mag, dashed line)
the $\delta^2$ function gets wider and `noisier', and 
for the strongest influence of microlensing
($\sigma_{\rm ML} =  0.1$ mag, thick solid line)
a secondary features appears. 
But in no case the distortion prohibits a clear  and correct
time delay determination,
the primary minimum is still clearly identifiable 
(as will be demonstrated by Wisotzki \& L\'opez, 2001, in preparation,
microlensing fluctuations during the period covered by our monitoring are of the
order of $0.07$ mag rms).

To make sure that this is not a chance observational
effect of this particular selected lag, we repeat this exercise
for an assumed shift of $-0.5$ years between the observed lightcurve
and its shifted copy, plus added `artificial microlensing' 
with $\sigma_{\mathrm{ML}} = 0.1$ mag. Again, the correct value
is clearly recovered in all realisations. This is particularly
convincing because a lag of 0.5 years is the `worst case scenario' 
with minimal overlap between the two lightcurves.
To summarize, moderate microlensing can be a cause of distortions
of the time delay determination function, but it is unlikely that
microlensing dominates it completely.

\begin{figure}[hbtp]
 \centering
 \epsfxsize=9 cm
 \rotatebox{0}{\epsffile{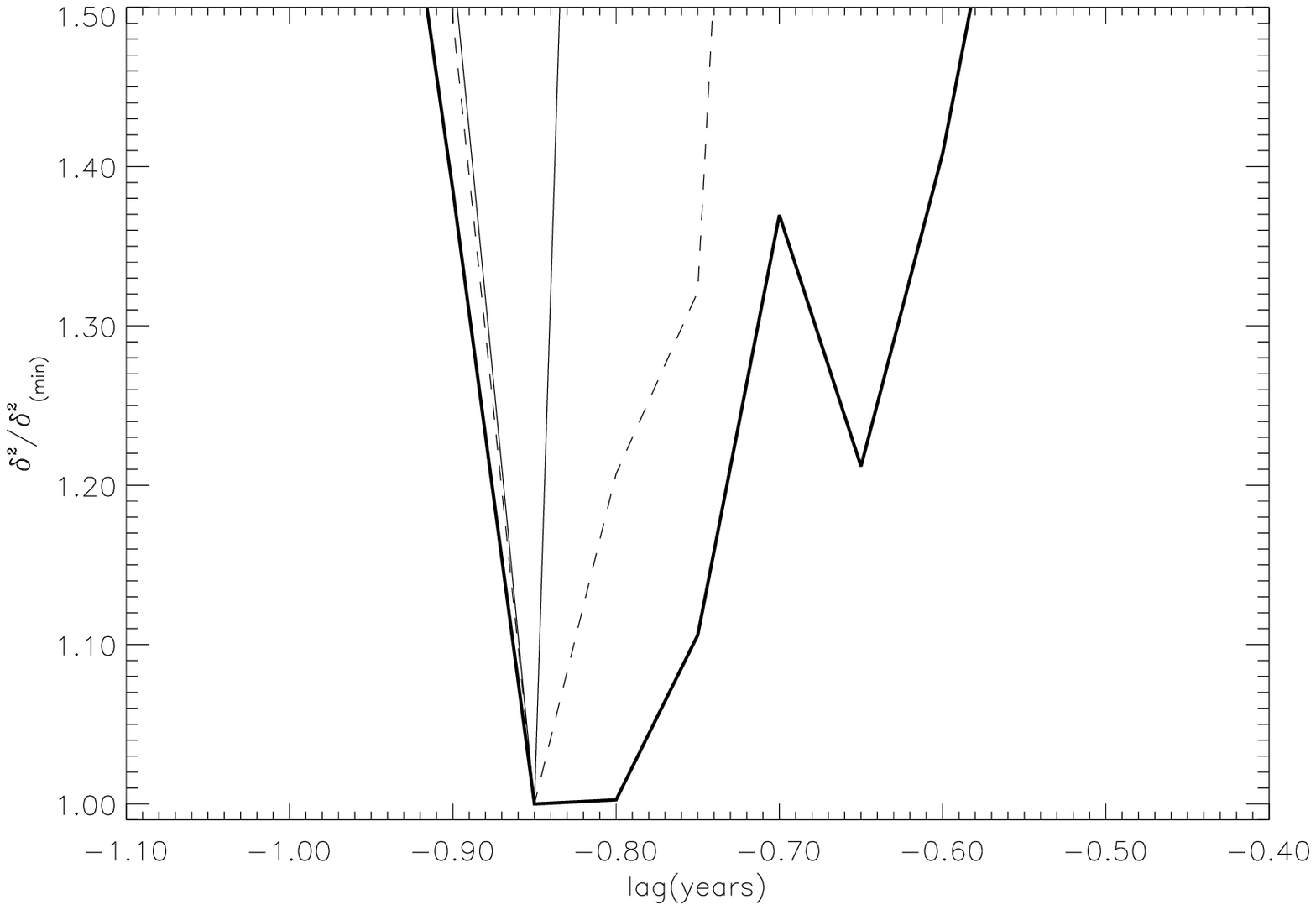}}
 \caption[]{We analyse the sampling effect in the $\delta^2$ technique. We use
 lightcurves B and B$'$, being B$'$ a copy of B shifted $0.85$ years and removing a
 number of points. \emph{Thin solid line}: we remove $2$ random points in the
 component B$'$. \emph{Dashed line}: when removing $4$ random points, it appears
 secondary structure in the $\delta^2$ function. \emph{Thick solid line}: if $3$
 selected points are remove, the $\delta^2$ normalized function is very similar
 to the one computed with lightcurves A and B (see Fig. \ref{Fig11}).}
 \label{Fig15}
\end{figure}
\subsubsection{Sampling}
In order to study the effect of sampling on the shape of the
$\delta^2$ function, we proceed as follows: 
again, we consider the lightcurve of component B and an identical
copy of it shifted by $0.85$ years, lightcurve B$'$. 
Now we remove some points from lightcurve B$'$. 
Resulting  $\delta^2$ functions are shown 
in Fig. \ref{Fig15} for three cases.
The thin solid line is  a case in which $2$ random points have been
removed from B.
The minimum of the $\delta^2$ 
normalized function is still well defined, with no secondary structure. 
For the dashed curve in Fig. \ref{Fig15}, $4$ random points were taken
away.
The shape of the function is distorted and a secondary dip appears. 
For the thick solid line, $3$ adjacent hand-picked points
(epochs $1997.12$, $1997.21$, $1997.27$) were excluded. 
Surprisingly, although all the remaining data points have 
identical spacing in B$'$ as in B, the removal of the 3 points
causes a secondary minimum  in the 
$\delta^2$ function,
which is very similar to the one obtained for the real
data, using the observed lightcurves A and B (Fig. \ref{Fig11}).
This case is very illustrative:  it suggest that the
sampling alone could be responsible for the secondary minimum
found in the real data (Fig. \ref{Fig11}).
This effect certainly deserves more study. From this preliminary 
analysis it appears that better and denser sampling of quasar
lightcurves could be much more important for time delay
studies than fewer data points with higher photometric precision.

As above, we also want to check whether the particular
value of the time lag plays an important role, and we again
repeat the simulation exercise with an assumed lag of $-0.5$ years,
and 4 randomly selected points removed. The result is again
$\Delta t_{B-B'} = -0.5$ years, recovering the assumed lag
in all cases.

\subsubsection{Summary of microlensing/sampling effects}
Summarizing, we can state that both microlensing and sampling differences
affect the shape of the time delay determination function. However,
moderate microlensing will have only small effects on these curves,
whereas moderate (and unavoidable!) differences in the sampling
for the two lightcurves can easily introduce effects like secondary
minima.  
The primary minimum of the $\delta^2$ method in all cases considered was
still clearly representing the actual value of the time delay.
Applied to HE~1104$-$1805, this means that most likely
microlensing does not affect much the time delay determination, 
the features in the time delay determination function can
be easily explained by the sampling differences, and
the primary minimum appears to be a good representation of
the real time delay.

\subsection{Implications for $H_0$ determination}
If one wants to use the time delay to estimate the Hubble
parameter $H_0$, one needs to know the geometry and mass distribution of the 
system. Accurate astrometry is available from HST images presented by \cite{Lehar00}. 
There are also several models for the lens in the literature. In W98,  
two models are described: a singular isothermal sphere with external shear and 
a singular isothermal ellipsoid without external shear. The first model is 
similar to \cite{Remy98} and \cite{Lehar00}. \cite{Courbin00} also present two 
models: a singular ellipsoid without external shear and a singular isothermal 
ellipsoid plus an extended component representing a galaxy cluster centered on 
the lens galaxy.

The redshift of the lens in this system has been establish by \cite{Lidman00} to
be $z_d=0.729$. Note that HE~1104$-$1805 is somehow atypical, in the sense that the
brightest component is closer to the lens galaxy. We use the most recents models 
by the CASTLES group (Leh\'ar et al.\ 2000), described by a singular isothermal 
ellipsoid (SIE) and a 
constant mass-to-light ratio plus shear model ($M/L+\gamma$). The derived 
value for the Hubble constant using the first model (SIE) is 
$H_0=48\pm4~\mathrm{km}~\mathrm{s}^{-1}~\mathrm{Mpc}^{-1}$ with $2\sigma$ 
confidence level. A ($M/L+\gamma$) model gives 
$H_0=62\pm4~\mathrm{km}~\mathrm{s}^{-1}~\mathrm{Mpc}^{-1}$ ($2\sigma$), 
both for $\Omega_0=1$. 
The formal uncertainty in these values are very low, due to the low formal 
uncertainties both in the time delay estimation and in the models. Nevertheless, 
the mass distribution is not well constrained, since a sequence of models can 
fit the images positions (Zhao \& Pronk 2000). We note that other models 
in \cite{Lehar00} will give very different results for $H_0$, but we did not use
them because no error estimate was reported for them. Moreover, the angular 
separation is big enough to expect an additional contribution to 
the potential from a group or cluster of galaxies 
(Mu\~noz 2001, priv. comm.).

\section{Conclusions}
We have shown that the existing data allow us to constrain the time delay of 
HE~1104$-$1805 with high confidence between 0.8 and 0.9 years,
slightly higher than the one available previous estimate. We have demonstrated
that the six different techniques employed in this study were not equally suited
for the available dataset. In
fact, this case study has demonstrated that a very careful analysis of each
technique is needed when applying it to a certain set of observations. Such an analysis
becomes even more important in the case of poorly sample lightcurves. In this sense, 
the $\delta^2$ technique showed the best behaviour against the
poor sampling: unless the lack of information due to sampling is so severe that it
prevents the determination of a well defined DAC and DCC, the minimum of the $\delta^2$ 
function will be a robust estimator for the time delay.

Our improved time delay estimate yields a value of the Hubble parameter which now depends
mostly on the uncertainties of the mass model. The degeneracies inherent to a simple
2-image lens system such as HE~1104$-$1805 currently preclude to derive very tight limits
on $H_0$. We note, however, that there are prospects to improve the constraints on the
model e.g.\ by using the lensed arclet features visible from the QSO host galaxy. Even
now, there seems to be a remarkable trend in favour of a relatively low value of $H_0$,
consistent with other recent lensing-based estimates (Schechter 2000).

\begin{acknowledgements}
We are especially grateful to Dr.L.J.Goicoechea (Universidad de Cantabria, Spain) 
for indicating us the best way of applying the $\delta^2$ technique and for many
comments to a first version of this paper. We also thank 
Dr.L.Tenorio (MINES, USA) for his useful comments on Monte Carlo and bootstrap 
simulations and Dr.J.A.Mu\~noz (IAC, Spain) for enlightening our understanding of lens
modeling.
\end{acknowledgements}
{}

\end{document}